\documentclass[aps,twocolumn,superscriptaddress,altaffilletter,lengthcheck, tightenlines,showpacs,showkeys]{revtex4}
\usepackage[colorlinks=true, pdfstartview=FitV, linkcolor=blue, citecolor=red, urlcolor=magenta, breaklinks=true]{hyperref}
\hypersetup{colorlinks,citecolor=blue,filecolor=magenta,linkcolor=magenta,urlcolor=cyan,pdftex}

\newcommand{\ben}{\begin{eqnarray}}
\newcommand{\een}{\end{eqnarray}}
\newcommand{\be}{\begin{equation}}
\newcommand{\ee}{\end{equation}}
\newcommand{\ba}{\begin{eqnarray}}
\newcommand{\ea}{\end{eqnarray}}

\newcommand{\bn}{\begin{equation}\label}

\usepackage[dvipdf]{epsfig}
\usepackage{color}

\begin{document}

\title{Relativistic Bose--Einstein condensates thin-shell wormholes}
%\title{ Evolving wormholes supported by modified Chaplygin gas}

%\title{Wormholes with relativistic Bose--Einstein condensates}

\author{M. G. Richarte}\email{martin@df.uba.ar}
\affiliation{Departamento de F\'isica, Universidade Federal do Paran\'a, Caixa Postal 19044, 81531-990 Curitiba, Brazil}
\affiliation{Departamento de F\'isica, Facultad de Ciencias Exactas y Naturales,
Universidad de Buenos Aires, Ciudad Universitaria 1428, Pabell\'on I,  Buenos Aires, Argentina}
\author{I. G. Salako}\email{inessalako@gmail.com}
\affiliation{D\'epartament de Physique, Universit\'e Nationale d'Agriculture, 01 BP 55 Porto-Novo, Benin}
\affiliation{Institut de Math\'ematiques et de Sciences Physiques, Universit\'e  d'Abomey-calavi Porto-Novo, 01 BP 613, Porto-Novo, Benin}
\author{J. P.  Morais Gra\c{c}a}\email{jpmorais@gmail.com}
\affiliation{Departamento de F\'isica, Universidade Federal do Paran\'a, Caixa Postal 19044, 81531-990 Curitiba, Brazil}
\author{H. Moradpour}\email{h.moradpour@riaam.ac.ir}
\affiliation{Research Institute for astronomy and Astrophysics of Maragha, Maragha 55134-441, Iran}
\author{Ali \"{O}vg\"{u}n}\email{ali.ovgun@pucv.cl}
\affiliation{Instituto de F\'{\i}sica, Pontificia Universidad Cat\'olica de Valpara\'{\i}so, Casilla 4950, Valpara\'{\i}so, Chile}
\affiliation{Physics Department, Eastern Mediterranean University, Famagusta,
Northern Cyprus}
\bibliographystyle{plain}

\begin{abstract}
We construct traversable thin-shell wormholes which are asymptotically Ads/dS  applying the cut and paste procedure for the case of an acoustic metric created by a relativistic Bose-Einstein  condensate. We examine several definitions of the flare-out condition along with the violation or not of the energy conditions  for such relativistic geometries. Under reasonable assumptions about the equation of state of the matter located at the shell,  we concentrate on the mechanical stability of  wormholes under radial perturbation preserving the original spherical symmetry.  To do so, we consider  linearized perturbations around  static solutions. We obtain that  dS acoustic wormholes  remain stable under radial perturbations as long as they have small radius; such wormholes with finite radius do not violate the strong/null energy condition. Besides, we show  that stable Ads wormhole satisfy some of the energy conditions whereas  unstable Ads wormhole with large radii violate them. 

\end{abstract}
\vskip 1cm

%\keywords{ modified Chaplygin gas, evolving-wormholes, exotic matter, energy conditions}
%\pacs{98.80.-k, 98.80.Jk}

\date{\today}
\maketitle

\section{Introduction}
 
One of the most fascinating outcomes of Einstein's theory of gravity concerns to  the existence of wormhole-like  geometries. Such hypothetical tunnels
connect  widely separated regions of spacetime from where in-going causal curves can pass through and become out-going on the other side \cite{morris1}.
Since the seminal work of  Morris and Thorne based on classical traversable wormholes \cite{morris1},  the theoretical progress on traversable wormholes 
was mainly motivated by  the possibility of constructing time machines \cite{morris2} by taking into account relative motions of  two wormhole's mouths, or equivalently the gravitational redshifts at the mouths due to external gravitational fields might produce closed time-like curves  \cite{morris2}. However,  one issue  related to wormhole physics is that within general relativity static wormholes are supported by exotic matter, which implies that the weak energy condition (WEC)  is violated \cite{visser1}, \cite{visser2}. Nevertheless,  some efforts were devoted to reduce the amount of exotic matter  as much as it is physically possible \cite{visser2}.  For instance, Visser \emph{et al.} developed a suitable measure for quantifying this notion and demonstrated that traversable wormholes can be sustained by an arbitrarily small amount of exotic matter \cite{visser3}. In the case of dynamical Lorentzian wormholes featured  by an overall time dependent conformal factor was possible  to achieve  configurations where WEC is not violated \cite{ancho1}. In fact,  
nonstatic Lorentzian wormholes  which evolve from a dS phase towards a FRW spacetime  not need to be threatened by  exotic matter \cite{ancho2}, \cite{ancho3}.  Besides, any attempt to construct thin-shell
wormholes requires the use of the cut-and-paste procedure \cite{visser1}  and works with the junction conditions associated
with the gravity theory under study \cite{cj0}, \cite{cj1}.   Spherically thin-shell wormholes within the context of general relativity were built, and it was found that, in most of the cases, the
wormholes are supported by exotic matter also, violating the energy conditions \cite{visser1}, \cite{visser2}.

It has been realized recently  that  starting from a relativistic Bose-Einstein condensate (BEC) \cite{libe1}, \cite{libe2}, \cite{libe3} one could (partially) reconstruct the notion of a  curved manifold along with the  general relativity scheme  from  first principles by means of  the emergent gravity  paradigm \cite{eg1}, \cite{eg2}, \cite{eg3}, \cite{eg4}.  BEC are  described by means of a self-interacting complex scalar field  coupled to an external potential. In doing so,   the dynamic of  condensed boson is governed by a non-linear  Klein-Gordon equation  and it leads to Gross-Pitaevskii equation  in the non-relativistic limit \cite{libe2}. Splitting the scalar field as a (classical) condensate field plus a quantum perturbation and using a Madelung representation for the mean (condensate) field,  one can prove that  the dynamic of phonon perturbation is governed by a modified Klein-Gordon equation \cite{libe2}. At the linear level,  the phonon propagates as a massless scalar field in the geometry  created by the condensate field. So, one can read off the effective (acoustic) metric created by the condensate from the wave equation satisfied by the phonons. Naturally, the emergent geometry associated with the effective metric created by the  condensate only involves the classical density of the condensate, its four-velocity  and a function which encodes its interaction strength \cite{libe2}, \cite{libe3}. Another interesting fact is that the emergent geometry is disformally related to the Minkowski background, but most importantly, such effective metric can be mapped into Schwarzschild-AdS and dS black holes, up to a conformal factor, by choosing  the normalized velocity profile properly \cite{map}.  Even better,   3-dimensional planar AdS black holes can be exactly obtained from  a non-relativistic BEC. However, higher dimensional black holes are conformally mapped
into acoustic geometries \cite{map}.

The plan of the paper is the following: we first consider  the  main ingredient of  the Bose-Einstein condensate in both relativistic and non-
relativistic regime and how  an effective acoustic metric associated with the classical condensate emerges (Sec. II). In Sec. III, we address the construction 
of thin-shell wormholes using the cut-and-paste procedure and  discuss the fulfillment of the flare-out condition for such configurations and the analysis of the energy conditions amongst other topics. Sec. III is devoted to the analysis of  the stability of the wormhole throat by choosing an equation of state for the matter located at the wormhole throat. In Sec. IV, the conclusion are stated. The signature of the metric is $(-, +,+,+)$.

\section{Relativistic BEC and Emergent gravity}
We start with the notion of BEC and its connection with the emergent gravity paradigm \cite{libe2}, \cite{libe3}. The  analogue gravity formalism relies on the idea that  after  the equations of fluid dynamics have been linearized under appropriate
conditions, the perturbations are accommodated as quasiparticles which obey a relativistic equation of motion in
a curved spacetime. In such scheme,  the emergent geometry (or  acoustic metric) is created by the condensate field  whereas the phonons propagate as  massless particles in the aforesaid background. To show so, let us consider  a massive complex scalar field with a self-interacting potential, ${\cal U}_{\rm s}$, in the presence of an external field, $ V_{\rm ext}$.  Its Lagrangian reads
\begin{eqnarray} \label{la}
{\cal L}= -\eta^{\mu\nu}\partial_{\mu}\phi^{\star} \partial_{\nu}\phi -\Big(\frac{m^{2}c^{2}}{\hbar^{2}}+ V_{\rm ext}  \Big)|\phi|^{2} -{\cal U}_{\rm s}(|\phi|, \lambda_{\rm i}),
\end{eqnarray}
where $m$ stands for the boson's mass,  $c$ is the speed of light, $\hbar$ is the  Planck's constant,  $\lambda_{\rm i}$ denotes  some coupling constants,  and    $\eta^{\mu\nu}$ is the Minkowski tensor. An interesting fact regarding (\ref{la}) is that  it remains invariant under  the global $U (1)$
symmetry. The  global charge is the difference between  the particle number of the ensemble composed  $N$ bosons and  the particle number for the anti-boson ensemble $\bar{N}$, then  
the Noether's theorem enables us to determine its associated (conserved) current, i.e. $J^{\mu}=i(\phi^{\star}\partial^{\mu}\phi-\phi\partial^{\mu}\phi^{\star})$.
We emphasize that there is an integral expression for the conserved number density $n$ in terms of its critical temperature $T_{c}$ (cf.  \cite{libe2}) when the ensemble is composed of  non-interacting particles characterized by ${\cal U}_{\rm s}=V_{\rm ext}=0$.  Consequently, the behavior of the number density with the  critical temperature in the ultra-relativistic limit and non-relativistic limit can be  extracted from that formula (for a nice review  see \cite{libe2} and references therein). Interestingly enough,  for temperatures below the critical one, the dynamic of the relativistic boson condensate is governed by a modified Klein-Gordon equation:
\begin{eqnarray}
\label{conmkg}
\eta^{\mu\nu}\partial_{\mu} \partial_{\nu}\phi -\Big(\frac{m^{2}c^{2}}{\hbar^{2}}+ V_{\rm ext}  \Big)\phi -{\cal U}'_{\rm s}\phi=0.
\end{eqnarray} 
For $T\ll T_{c}$ the thermal effects are neglected,  so one  decomposes the scalar field as a BEC field (ground state) plus its excitations by using the following parametrization, $\phi= \varphi (1+ \psi)$ being $\varphi= \left\langle \phi \right\rangle$ the condesated part. The condensate field satisfies the same modified K-G equation (\ref{conmkg}). Replacing  the previous ansatz in (\ref{conmkg}), we arrived at the master  equations which obey the condensate part plus the wave-like equation for the phonons:
\begin{eqnarray}
\label{rho}
\frac{\hbar^2}{m^2} \Big(V_{\rm ext}+ {\cal U}'_{\rm s}(\rho, \lambda_{\rm i})- \frac{\eta^{\mu\nu}\partial_{\mu}\partial_{\nu}\sqrt{\rho}}{\sqrt{\rho}}\Big)=-u^{\mu}u_{\mu}-c^{2},
\end{eqnarray} 
\begin{eqnarray}
\label{psi}
\left[{\cal D}c^{-1}_{0}{\cal D}^{\dagger} -\frac{\hbar^{2}}{\rho} \eta^{\mu\nu}\partial_{\mu}\rho\partial_{\nu}\right] \psi=0,
\end{eqnarray} 
where we have employed a Born-like identification for the four-velocity, namely $u^{\mu}\equiv (\hbar/m)\partial^{\mu}\theta$. We also defined the operator ${\cal D}=i\hbar u^{\mu}\partial_{\mu}+ T_{\rm gk}$ being  $T_{\rm gk}=-(\hbar^{2}/2m)[\eta^{\mu\nu}\partial_{\mu}\partial_{\nu}+ \eta^{\mu\nu}\partial_{\mu} \ln \rho\partial_{\nu}] $ (generalized kinetic operator). The curvature of the self-interacting boson potential is encoded in the strength interaction function called $c^{2}_{0}=(\hbar^{2}/2m)\rho {\cal U}''_{\rm s}$. Notice that the conserved current satisfies the continuity equation: $\partial_{\mu}J^{\mu}=\partial_{\mu}(\rho u^{\mu})=0$. 

To analyze further how the acoustic metric emerges for the quantum perturbation it might be useful to introduce a series of well accepted approximations. We are going to work within the phononic (infrared relativistic) regime, which implies that Eq. (\ref{psi}) emulates the dispersion relation of phonons, $\omega= c_{s}k$, where the squared speed of sound is $c^{2}_{s}\equiv [(cc_{0}/u^{0})^2/(1+ (c_{0}/u^{0})^2)]$.  Notice that such assumption is equivalent to the low-momentum  condition derived in Ref. \cite{libe2}. Second, all relevant background quantities will vary slowly over scale comparable with the perturbation wavelength, thus $|\partial_{t}X/X|\ll w$ with $X=\{\rho, u^{\mu}, c_{0} \}$.  The point is to have a situation where the generalized  kinetic operator can be neglected so we can guarantee that the low energy limit holds. Then, neglecting the quantum contribution encoded in  $T_{\rm gk}$  and using the $\partial_{\mu}J^{\mu}=0$, Eq. (\ref{psi}) can be written as
\begin{eqnarray}
\label{psi2}
\partial_{\mu}\big(\gamma^{\mu\nu}\partial_{\nu} \psi\big)=0.
\end{eqnarray}
Here we introduced the auxiliary metric tensor $\gamma^{\mu\nu}\equiv (\rho/c^{2}_{0})[-u^{\mu}u^{\nu}+c^{2}_{0}\eta^{\mu\nu}]$. In order to make the  D'Alembertian operator appears in (\ref{psi2}) it is mandatory  that to perform an identification such as $\gamma^{\mu\nu}\equiv \sqrt{-g}g^{\mu\nu}$ with $\sqrt{-g}=\rho^{2}[1-u^{\nu}u_{\nu}/c^{2}_{0}]^{1/2}$ so the inverse metric is given by  $g^{\mu\nu}= \gamma^{\mu\nu}/\sqrt{-g}$. Calculating the inverse of $g^{\mu\nu}$ we obtain the effective or acoustic metric for the phonon perturbations, that explicitly reads $g_{\mu\nu}=(\rho/\sqrt{1-u^{\beta}u_{\beta}/c^{2}_{0}})\big[\eta_{\mu\nu}\big(1-u^{\beta}u_{\beta}/c^{2}_{0}) + u^{\nu}u^{\mu}/c^{2}_{0}\big]$. In terms of the acoustic metric (\ref{psi2}), one finds that Eq. (\ref{psi2}) takes the traditional form of the wave equation for a massless scalar field (phonon) in curved spacetime: $\triangle \psi=(1/\sqrt{-q})\partial_{\mu}[ \sqrt{-g}g^{\mu\nu}\partial_{\nu}\psi]=0$. For practical purposes, we write the acoustic metric in a more useful form as follows
\begin{eqnarray}
\label{acou2}
g^{\rm acoustic}_{\mu\nu}=\Omega \big[\eta_{\mu\nu} + \big(1-\frac{c^{2}_{s}}{c^2}\big)\frac{v^{\mu}v^{\nu}}{c^2}\big],
\end{eqnarray}
where the conformal factor is defined as $\Omega=\rho c/c_{s}$ and the normalized four velocity field is $v^{\mu}\equiv c u^{\mu}/|u|$.  The shift introduced by the dyadic tensor $v^{\mu}v^{\nu}$ shows up that the acoustic metric  is disformally related with the Minkowski background. Besides, as it was done previously in Ref. \cite{map} one must write the above metric (\ref{acou2}) in ``minkowskian'' coordinates $(ct, x^{i})$ used in the lab system. The line element in that coordinates takes the next form
\begin{eqnarray}
\label{met1}
ds^{2}_{\rm ac}=\Omega \big[{\cal G}_{00}c^2dt^2+ 2{\cal G}_{0i}cdtdx^{i} + {\cal G}_{ij}dx^{i}dx^{j}].
\end{eqnarray}
The coefficients of the effective metric are listed below
\begin{eqnarray}
\label{coef1b}
{\cal G}_{00}=-1 + \xi \frac{v^{2}_{0}}{c^{2}}, ~ {\cal G}_{0i}=\frac{v_{0}v_{i}}{c^2}, ~{\cal G}_{ij}=\delta_{ij}+ \xi \frac{v_{i}v_{j}}{c^2}.
\end{eqnarray}
Here we introduced the parameter $\xi=1-c^{2}_{s}/c^{2}$. With the help of the normalization condition, $v^{2}=-c^2$, one can recast the above coefficients (\ref{coef1b}) by changing $v_{0}=\pm c \sqrt{1+ |v^{i}|^{2}/c^{2}}$.  Eq. (\ref{met1}) tells us that the non-relativistic limit is reached by imposing the regime of low velocity ($v\ll c$ plus $c_{s}\ll c$, that is, $\xi \simeq 1$) along with the weak interaction condition, namely $c_{0}\ll c$. 

Before embark us in the construction of thin-shell wormholes it is essential to prove that (\ref{met1}) can be recast in a suitable form, where it takes a diagonal form \cite{map} . To do so,   one must implement a change of coordinates given by 
\begin{eqnarray}
\label{change}
cdt=cd\eta\pm {\cal P}_{i}dx^{i},
\end{eqnarray}
where the field ${\cal P}_{i} \equiv\xi (v_{0}v_{i}/c^{2})/[-1+ \xi v^{2}_{0}/c^{2}]$. As it was noticed by Cropp \emph{et al.} \cite{map2}, the vector field must be obtained from a scalar function ($ {\cal P}_{i}=\nabla_{i} \phi$). This implies that an integrability condition should hold: $\epsilon_{ijk}\partial_{j} {\cal P}_{k}=0$.  Taking into account the latter facts, Eq .(\ref{change}) helps us to recast the metric in a new stationary form given by the line element:
\begin{eqnarray}
\label{met2}
ds^{2}_{\rm ac}=\Omega\left[ -(c^{2}_{s}-\xi |\bar{v}|^2)d\eta^2+ \big(\delta_{ij} + \frac{\xi v_{i}v_{j}}{c^{2}_{s}-\xi|\bar{v}|^2}\big)dx^{i}dx^{j}\right].~~~
\end{eqnarray}
An interesting point regarding (\ref{met2}) is that by choosing properly the velocity profile $v^{i}$, one can show that the above metric can be mapped into black holes which are  asymptotically AdS or dS at the spatial infinity \cite{map}, \cite{map2}.  To do so, we simply follow the recipe adopted in \cite{map} and \cite{map2}. Let us start by considering the metric associated with Ads/dS black holes in a spherical coordinate patch $(c\eta, r, \theta, \phi)$:
\begin{eqnarray}
\label{met3}
ds^{2}_{\rm bh}=f(r)c^{2}_{s}d\eta^2 + g(r) dr^{2} + r^{2}(d\theta^2+ {\rm sin}^2 \theta d \varphi^2),
\end{eqnarray}
where the metric coefficient are $f(r)=[g(r)]^{-1}=[1-(r_{0}/r)\pm (r/L)^{2}]$ with  $r_{0}$ a constant. The cosmological constant is  $\Lambda=L^{-2}$ and the $\pm$ denotes if the black hole is asymptotically Ads (+) or dS (-), respectively.  To demonstrate that the black hole metric (\ref{met3}) can be mapped into the relativistic effective metric of BEC (\ref{met2}), we propose that the velocity field must be spherically symmetric, and thus the only non-zero component is the radial one, $\bar{v}=v^{i}\delta^{r}_{i}$.  After having compared the $\eta-\eta$ component of the line elements (\ref{met2})-(\ref{met3}), we arrive at the profile of normalized velocity field, $v^{2}_{r}=(c^{2}_{s}/\xi)[(r_{0}/r)\mp (r/L)^{2}]$. The unnormalized vector field can be easily obtained by using the definition of $v^{\mu}$. As a consequence we obtain the nice relation $u_{r}=(v_{r}/v_{0})u_{0}$ or equivalently $u^{r}=[v^{r}/\sqrt{1+ v^{2}_{r}/c^{2}}](u_{0}/c)$. Taking into account all the mentioned facts, the unnormalized  radial flow is given by 
\begin{eqnarray}
\label{flow}
u^{r}=\frac{u_{0}c_{s}}{c\sqrt{\xi}} \frac{\sqrt{\frac{r_{0}}{r} \mp \frac{r^{2}}{L^2}}}{\sqrt{1+ \frac{c^{2}_{s}}{c^{2}\xi}\big(\frac{r_{0}}{r} \mp \frac{r^{2}}{L^2}\big)}}.
\end{eqnarray}
At this point, we use the continuity equation in spherical coordinates to reconstruct the density profile. Using that $\partial_{r}(r\rho u^{r})=0$, we find that $\rho= \rho_{0}/[r u^{r}]$ with $\rho_{0}$ a constant.  Some comments are in order. The above results show that it is possible to connect black holes metric with acoustic effective geometry associated with BEC up to a conformal factor, nevertheless, we would  like to stress that such relation can be extended at the level of the gravitational field equation. To be more precise, it was shown by Dey \emph {et al.} \cite{map} that the gravitational field equations of the relativistic BEC share some similarity with the Einstein-Fokker equation for a Nordstrom gravity (see the appendix A of ref. \cite{map}). Having said that, let us also mention that analogy is not completely due to the extra-conformal factor of the acoustic metric. Notice that in the case of asymptotically Ads background,  the velocity profile together with the density profile are well definite as long as the inequality $r^{3}<r_{0}L^2$ can be guaranteed.  Finally, the acoustic black hole metric can be recast as 
\begin{eqnarray}
\label{metfinal}
ds^{2}_{\rm acbh}=\Big[-{\cal F}d\eta^2+ {\cal G}dr^{2} + {\cal H}\big(d\theta^{2} + {\rm sin}^2 \theta d \varphi^2\Big)\Big], 
\end{eqnarray}  
where the metric coefficients are given by 
\begin{eqnarray}
\label{coeffinal}
{\cal F}= \Omega f(r) c^{2}_{s},~~ {\cal G}= \Omega [f(r)]^{-1}, ~~ {\cal H}=\Omega r^{2}.
\end{eqnarray}  
To end  with this section, let us just mention  that the acoustic black hole has particular regions which deserve some particular attention. For the acoustic Ads (+) solution we have that the spacetime is similar to  the Schwarzschild  black hole for small $r$, up to the conformal factor, and approaches to Ads space for large $r$.  The black  hole  exhibits an event horizon at $r=r_{+}$, corresponding to the largest root of $f(r=r_{+})=0$.  On the other hand, 
the dS case has different possibilities to examine. Let us begin by noting that the condition $f(r)=0$ is equivalent to $r^{3}-L^{2}r+L^{2}r_{0}=0$, which might have different kinds of roots depending on the relation between $L$ and $r_{0}$. The discriminant of the cubic equation, namely $\Delta=L^4[(r_{0}/2)^2 -(L^{2}/27)]$, is the responsible for the classification of the roots. For $r_{0}=2L/3\sqrt{3}$, all roots are real and one is a double root while in the case of $r_{0}>2L/3\sqrt{3}$ two roots are complex conjugate and one is real. For $r_{0}<2L/3\sqrt{3}$, the cubic equation only has real roots.   % has two important hypersurfaces, the inner horizon event (the smallest root of $f(r)=0$, say $r_{-}$) and the cosmological event associated with the largest root of $f(r)=0$; these roots are given by    
%\begin{eqnarray}
%\label{roo1}
%r_{\pm}=\frac{r_{0}\pm \sqrt{r^{2}_{0}+\frac{4}{L^2}}}{2}.
%\end{eqnarray} 

\section{Thin-shell construction for acoustic black hole}
\subsection{General Method}
In this section, we are going to make use of the cut-and-past procedure to derive the geometry associated with thin-shell wormholes. To begin with, we consider the acoustic black hole metric (\ref{metfinal}) in order to build a spherically symmetric  acoustic thin-shell wormhole. For the asymptotically Ads metric, we take two copies of the space-time and remove from each manifold  the four-dimensional regions which contain event horizons, so the restricted spacetime is described by 
\begin{equation}
{\cal M}_{Ads\pm}=\left\{x/r_{\pm}\leq a,a>r_{+}\right\}.
\end{equation} 
Note that $a$ is chosen to exclude possible singularities or horizons within the region ${\cal M}_{\pm}$.  Here, we see that for an asymptotically dS black hole, we must exclude the event horizon along with the cosmological horizon, namely we ended up with the patch ${\cal M}_{dS\pm}=\left\{x/a\geq r_{-} \cap  a\leq r_{+}  \right\}$. Hence, we are dealing  with wormholes with finite radii in the latter case.  The resulting manifolds have boundaries given by the time-like hypersurfaces, 
\begin{equation}
\Sigma_{\pm Ads}=\left\{x/r_{\pm} = a,a>r_{+}\right\}.
\end{equation}
In the dS case the boundaries are described by the surface $\Sigma_{\pm dS}=\left\{x/r_{\pm} = a, a \in (r_{-}, r_{+})\right\}$.
Then we identify these two time-like hypersurfaces to obtain a geodesically complete new manifold ${\cal {M}}={\cal {M}}^{+}\cup {\cal {M}}^{-}/\Sigma_{\pm}$,  with a matter shell at the surface $r=a$ where the throat of the wormhole is located. This manifold is constituted by two regions which are   asymptotically Ads/dS, respectively. To study this type of wormholes we apply the Darmois-Israel formalism \cite{cj0}, \cite{visser1} to the case of the acoustic metric given by BEC.  We should stress this procedure is possible because several authors pointed out that the gravitational field equations of the relativistic BEC share some similarity with the Einstein-Fokker equation for a Nordstrom gravity (cf. \cite{map}). Then, the  projected field equations have to be analogue  to   the Darmois-Israel formalism \cite{cj0}, \cite{visser1}. In any case, we can take such junction conditions as the effective ones as well.    
\begin{figure}[!h]
\begin{center}
\includegraphics[height=9cm, width=8cm]{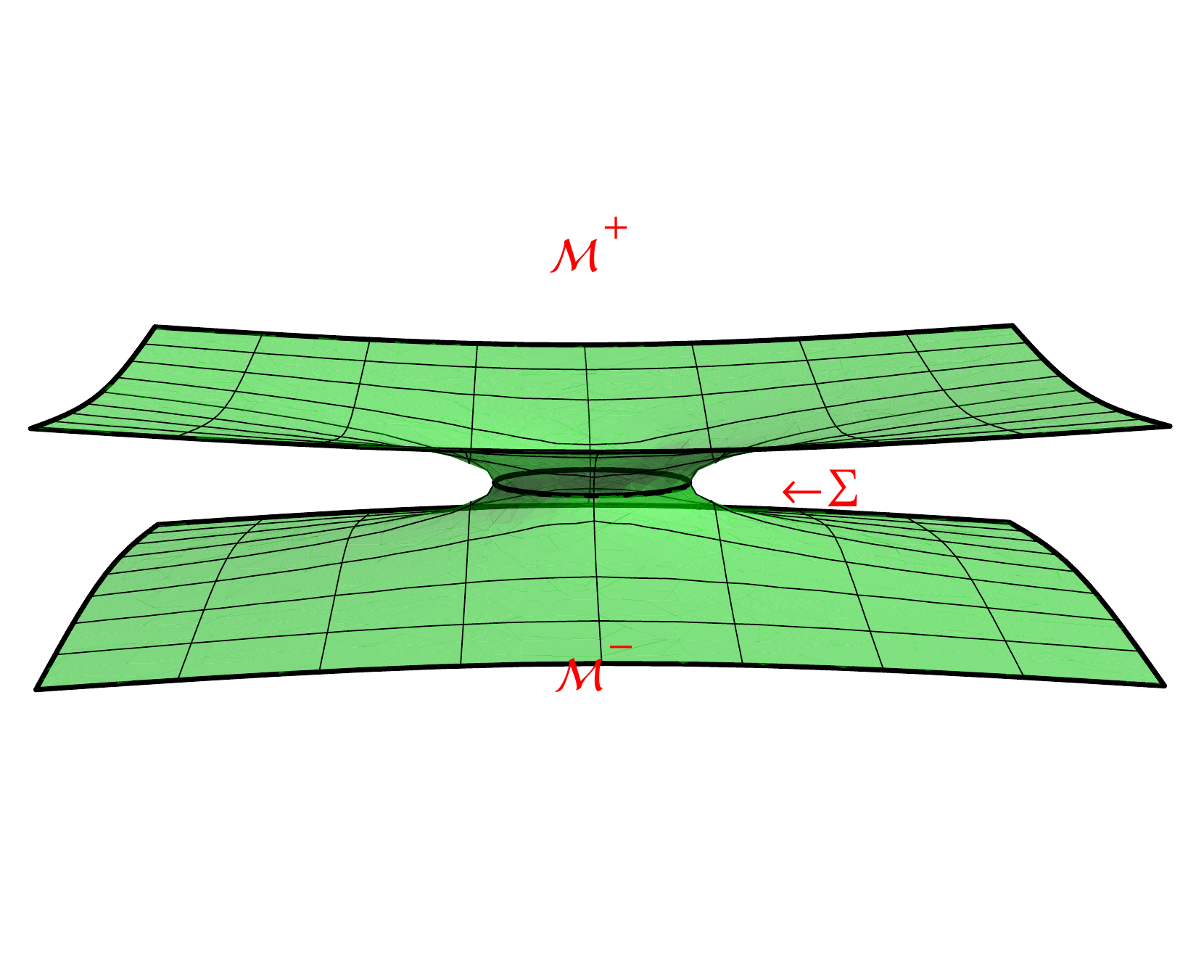}
\caption{Schematic representation for the wormhole geometry obtained after performing the cut and paste procedure. The shell on $\Sigma$ is located at the throat radius $r=a$.}
\end{center}
\end{figure}
We can introduce the coordinates $\xi^{a}=(\tau, \theta,\phi)$ in $\Sigma$, with  $\tau$ the proper time on the throat.  We will  focus  not necessarily in static configurations only, then the  boundary hypersurface reads:
\begin{equation}
\Sigma: {\cal H}(r, \tau)=r-a(\tau)=0.
\end{equation}
 
The  field equations projected on the shell $\Sigma$ are \cite{visser1}, 
\begin{equation}
\label{jc}
{\cal S}_{ab}=-\frac{1}{8\pi}\langle {\cal K}_{ab}-\gamma_{ab} {\cal K}\rangle,
\end{equation}
where the bracket $\left\langle .\right\rangle$ stands for  the jump of a given quantity across the  hypersurface $\Sigma$ and  $\gamma_{ab}$ is the induced metric on $\Sigma$. The extrinsic curvature ${\cal K}_{ab}$ is defined as follows:

\begin{equation}
{{\cal K}}^{\pm}_{ab}=-n^{\pm}_{A}\left(\frac{\partial^{2}X^{A}}{\partial\xi^{a}\partial\xi^{b}}+\Gamma^{A}_{B\,C}\frac{\partial X^{B}}{\partial\xi^{a}}\frac{\partial X^{C}}{\partial\xi^{b}}\right)_{r=a},
\end{equation} 
where $n^{\pm}_{A}=(n_{\eta}, n_{r}, 0,0)$ are the unit normals  to the surface $\Sigma$. In order to proceed one can write the intrinsic metric to $\Sigma$ as 
\begin{equation} 
\label{Throat}
ds^{2}_{\Sigma}=-d\tau^{2} + {\cal H}(a)(d\theta^{2}+\mbox{sin}^{2}\theta d\phi^{2}).
\end{equation}
The position of the junction surface is given by $X^{A}=(\eta(\tau), r(\tau), \theta, \phi)$ and the corresponding 4 velocity is $u^{A}=\big([\dot{\eta}(\tau),\dot{r}(\tau),0,0,0 \big)$,  where the dot stands for derivative with respect to $\tau$  and the surface $\Sigma$ is parametrized by giving $\eta=\eta(\tau)$ and $r=a(\tau)$. The unit normal to the shell may be determined by  the conditions $u^{A}n_{A}=0$ and $n^{B}n_{B}=1$. These requisites lead to the following expression: 
\begin{equation} 
\label{unit}
n^{\pm}_{A}=\epsilon\big(-\dot{a} [{\cal F}(a){\cal G}(a)]^{1/2},[{\cal G}(a)(1+{\cal G}(a)\dot{a}^{2})]^{1/2},0,0\big),
\end{equation}
where $\epsilon=\pm$ indicates if the normal is outward-pointing or inward-pointing, respectively.  We next compute the mixed components of 
the extrinsic curvature (second fundamental form)
\begin{eqnarray}
\label{KT} 
\left\langle {\cal K}^{\tau}_{~\tau} \right\rangle&=&\sqrt{\frac{{\cal G}(a)}{\ell}} \left[2\ddot{a}+ \frac{{\cal F}'(a)}{{\cal F}(a){\cal G}(a)} + \dot{a}^{2}\Big(\frac{{\cal F}'(a)}{{\cal F}(a)}+\frac{{\cal G}'(a)}{{\cal G}(a)} \Big)\right], ~~
\end{eqnarray}
\begin{eqnarray}
\label{KE} 
\left\langle {\cal K}^{\phi}_{~\phi}\right\rangle&=&\sqrt{\frac{\ell}{{\cal G}(a)}}\frac{{\cal H}'(a)}{{\cal H}(a)}=\left\langle {\cal K}^{\theta}_{~\theta}\right\rangle.
\end{eqnarray}
Here, $\ell(a, \dot{a})=1+{\cal G}(a)\dot{a}^{2}$ and the prime stands for derivative with respect $a$. Before analyzing the physical consequences of our model, we need first to determine the  energy-momentum for the matter located at the wormhole throat. To do so,  we  write down the  most general form of the stress-energy tensor on the shell which is compatible with the space-time symmetries 
 \begin{equation}
 \label{tem}
 {\cal S}^{a}_{~b}=~\mbox{diag}~(-\sigma, P_{\theta}, P_{\theta}).
\end{equation}
We want to determine the form of the energy density and tangential pressure in terms of the metric coefficients and the derivatives of the wormhole's throat. The way to achieve such goal is by combining the junction conditions (\ref{jc}), and  the mixed components of the second fundamental form (\ref{KE})-(\ref{KT}) along with the stress tensor (\ref{tem}).  After some algebraic manipulation, we obtain that the energy density  and the tangential pressure  can be recast as
\begin{eqnarray}
\label{sigma} 
\sigma=-\frac{1}{4\pi}\frac{{\cal H}'}{{\cal H}}\left(\frac{\ell}{{\cal G}}\right)^{1\over2},
\end{eqnarray}
\begin{eqnarray}
\label{pe} 
P_{\theta}= \frac{1}{8\pi}\left(\frac{{\cal G}}{\ell}\right)^{1\over2} \left[2\ddot{a}+ \frac{{\cal F}'}{{\cal F}{\cal G}} +\frac{{\cal H}'}{{\cal H}{\cal G}}  + \dot{a}^{2}\Big(\frac{{\cal F}'}{{\cal F}}+\frac{{\cal G}'}{{\cal G}}+ \frac{{\cal H}'}{{\cal H}} \Big)\right].
\end{eqnarray} 
The conservation equation for the wormhole's throat \cite{transp} reads
\begin{eqnarray}
\label{tra1} 
\nabla_{a}{\cal S}^{a}_{b}=-\left\langle T_{\alpha\beta} \frac{X^{\alpha}}{\partial \xi^{b}} n^{\beta} \right\rangle,
\end{eqnarray}
where  the operator $\nabla$ denotes the covariant derivative with respect to the induced metric and $T_{\alpha\beta}$ indicates the energy-momentum associated with the bulk matter. Eq. (\ref{tra1}) tells us that there is a transfer of energy between the shell located at $\Sigma$ and the bulk ${\cal M}$, 
\begin{eqnarray}
\label{tra2} 
\dot{[\sigma {\cal A}]} + p\dot{{\cal A}}= \delta {\cal Q}=-\frac{\sigma}{2} {\cal A}\dot{a}\Big(\frac{{\cal F}'}{{\cal F}}+\frac{{\cal G}'}{{\cal G}}+ \frac{{\cal H}'}{{\cal H}} -2\frac{{\cal H}''}{{\cal H}'} \Big).
\end{eqnarray}
Note that $\delta {\cal Q}$ vanishes for solutions with high spherical symmetry which fulfill the next relationships: i-${\cal G}={\cal F}^{-1}$ and ii-${\cal H}=r^{2}$. Nevertheless, this is not our case provided the global factor $\Omega$ appears in all the coefficients of the metric, which in turn  spoils all the possible cancellations. To be more precise, we obtain that $({\cal F}'/{\cal F})+({\cal G}'/{\cal G})=2\Omega'(a)/\Omega(a)$ along with $({\cal H}'/{\cal H}-2{\cal H}''/{\cal H}')=2[\Omega''a^{2}+4\Omega' a+2\Omega]/(\Omega'a^{2}+ 2a\Omega)$.  

\subsection{Flare-out condition and transversability}
One of the main ingredients of the wormholes construction refers to the flare-out condition \cite{fo1}, \cite{fo2}, \cite{broni}, \cite{habi}. Physically speaking, the aforesaid condition is equivalent to have a minimal surface represented by the wormhole throat which ensure that an observer can pass through.  Wormhole  geometries such as the one described by  Eq. (\ref{Throat})  admit different variants of throat definitions \cite{broni},  \cite{habi}. Let us start by mentioning the simplest one. To do so, we consider a local patch on $\Sigma\simeq \Re\times \Sigma_{\tau}$ given by $(\tau, y^{i})$ where $i=\theta, \phi$ such that the induced metric on $\Sigma$ can be recast as $h_{ab}={\rm diag}(-1, {\cal H}(a), {\cal H}(a) \mbox{sin}^{2}\theta)$ and the line element is given by Eq. (\ref{Throat}). In fact, we are going to discuss properties of the spatial section $\Sigma_{\tau}$ so  we must employ the induced metric on it, that is, $\gamma_{ij}={\rm diag}({\cal H}(a), {\cal H}(a) \mbox{sin}^{2}\theta)$. On the one hand,  a Morris-Thorne wormhole throat, at a given static time, is a minimal surface in the static hypersurface, i.e. locally minimizing area among surfaces in the hypersurface. In another words, a traversable wormhole throat should be considered as a two dimensional surface where  (i)- $\delta {\cal A} (\Sigma_{\tau})=0$ and (ii)-$\delta^{2} {\cal A} (\Sigma_{\tau})\geq 0$ are  met \cite{morris2}.
On the other hand, Hochberg and  Visser introduced the idea  that the throat of a traversable wormhole should be considered as a two dimensional surface $\Sigma_{\tau}$ with one important property, namely,  the ``flaring-out'' condition only  expresses strict minimality \cite{fo2}. The former conditions can be guaranteed when ${\rm Tr}({\cal K}^{i}_{~j})=0$ and $\partial_{n}({\rm Tr}({\cal K}^{i}_{~j}))\leq 0$, where $\partial_{n}$ stands for the directional derivative along the normal.  Nevertheless, the null trace requisite does not hold for thin-shell wormholes \cite{habi}. As it was pointed by  H. Mazharimousavi and M. Halilsoy the notion of thin-shell wormhole with matter located at its throat leads to ${\rm Tr}({\cal K}^{i}_{~j})=2{\cal K}^{\theta}_{~\theta}=-\sigma \neq 0$ \cite{habi}. One way to generalize the notion of a wormhole's throat is by  retaining the idea of minimal surface ($\partial_{n}({\rm Tr}({\cal K}^{i}_{~j}))\leq 0$)  and admit the possibility of  having  ${\rm Tr}({\cal K}^{i}_{~j})<0$ or  ${\rm Tr}({\cal K}^{i}_{~j})>0$ \cite{habi}. Let us illustrate what happens with our model in the case of a static thin-shell wormhole, namely  $\dot{a}=\ddot{a}=0$ and therefore $\ell=1$.  The trace condition yields
 \begin{equation}
 \label{trace}
{\rm Tr}({\cal K}^{i}_{~j})=\frac{2{\cal H}'(a)}{{\cal H}(a)\sqrt{{\cal G}(a)}}.
\end{equation}
Hence, the sign of  (\ref{trace}) is determined by the sign of  ${\cal H}'(a)$ provided that ${\cal H}$ is positive definite. Let us examine the condition (\ref{trace}) for acoustic wormholes which are topologically equivalent to Ads/dS at spatial infinity.  In the Ads branch, the original manifold only has one horizon which is located at $r_{+}=L$ by taking $r_{0}=2L$ and $L=1$ without loss of generality.  We obtain that ${\rm Tr}({\cal K}^{i}_{~j})>0$ for  large and small wormhole radii. We notice that such configurations can be  relativistic or non-relativistic ones provided that the trace condition remains positive for $\xi\in (0,1)$ (see  Fig. 2). It should be stressed that the relativistic nature or not of the acoustic wormholes is encoded in the allowed values taken by $\xi$.
\begin{figure}[!h] \label{fig2}
\begin{center}
\includegraphics[height=9cm, width=8cm]{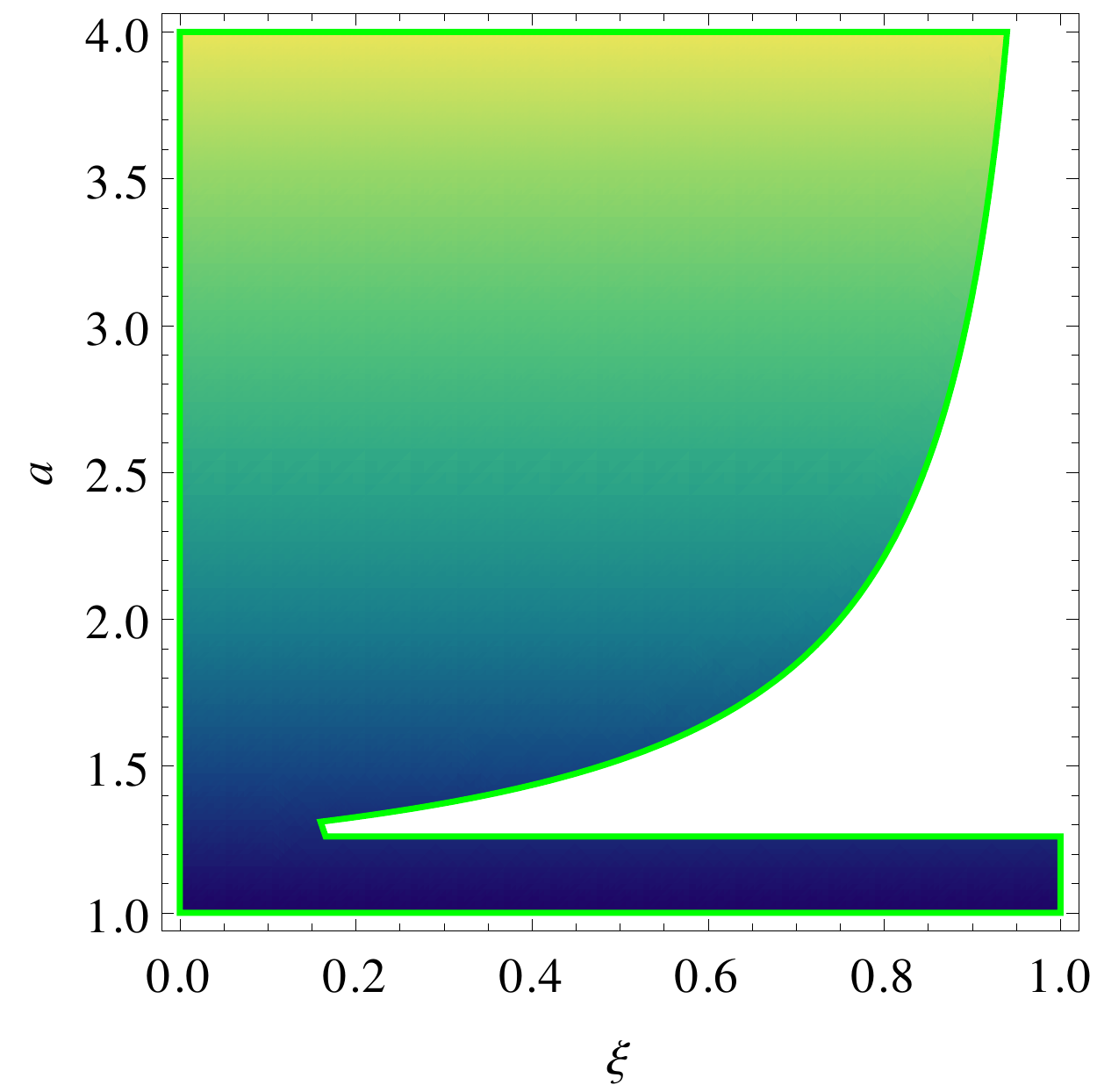}
\caption{It is plotted the fulfillment of the flare-out condition ${\rm Tr}({\cal K}_{ab})>0$ (shaded region) in terms of the wormhole radius  when $\xi \in (0,1)$. This case corresponds to  acoustic wormholes which are asymptotically Ads.}
\end{center}
\end{figure}
In the case of dS acoustic wormholes we find a similar situation regarding the sign of the trace condition. For $L=1$ and $r_{0}=1/4$, the original manifold has an inner horizon and the cosmological horizon placed at $r_{-}\simeq 0.26$ and $r_{+} \simeq 0.83$, respectively. When the wormhole radius $a \in (r_{-}, r_{+})$ and $\xi \in (0,1)$, we arrive at the condition ${\rm Tr}({\cal K}_{ij})>0$ , which is equivalent to state that the static configurations are supported by negative energy density. When the original manifold has a double horizon at $r_{\pm}=1/\sqrt{3}$ in the case of $L=1$ and $r_{0}=2/3\sqrt{3}$, the trace condition cannot be employed to characterize the transversavility of the configuration; the main reason is that the factor $\sqrt{G(a)}$ is ill definite because $G(a)<0$ in the aforesaid case. The same happens for the case of a geometry without horizons, corresponding to a situation in which $r_{0}=L/2$ for $L=1$. Nevertheless, we can apply less restrictive flare-out conditions which only require the positivity of ${\cal H}'(a)$. To show so, we must link the previous results with the applicability of standard flare-out condition, and therefore we must calculate the area of the surface $\Sigma_{\tau}$, thus ${\cal A}(\Sigma_{\tau})=4\pi {\cal H}(a)$. It is  straightforward to show that a minimal surface is obtained when ${\cal A}'(\Sigma_{\tau})>0$, which also implies the positivity of ${\cal H}'(a)$. In fact, it can be shown that this provides  a bound on the way the conformal factor $\Omega(a)$ can grow:
 \begin{equation}
 \label{trace2}
\frac{\partial \ln [ a^{2}\Omega (a)] }{\partial a}> 0.
\end{equation}  
Further,  one can consider another less restrictive flare-out condition \cite{broni} by using the perimeter of $\Sigma$ given by ${\cal P}=2\pi \sqrt{{\cal H}(a)}$ as a useful measure of the openness of the  wormhole's throat.  It is clear that the perimeter must be an increasing function of the wormhole's radius, so one must ensure that $\partial_{a}  \ln {\cal P}=\partial_{a}[\ln {{\cal H}}^{1/2}]>0$. The  minimality areal [${\cal A}'(\Sigma_{\tau})>0$] along with the  perimeter condition [${\cal P}' (\Sigma_{\tau})>0$]   are both satisfied provided  the inequality ${\cal H}'(a)>0$ holds for  the Ads/dS acoustic wormholes. %It should emphasized that the areal and perimeter conditions  % large space parameter when $a \in (r_{+}, \infty) $ and $\xi \in (0,1)$for $\xi\in(0,1)$.

%%Graficar ambas condiciones y ver donde se intersecan
%Consider a spatial section of the spacetime represented by the conditions $\eta=r=consts.$, the area is parametized as ${\cal A}=4\pi {\cal H }(a)$. Hence, the condition for a minimal surface is that ${\cal A}'(a)>0$ for all $a>r_{+}$ in the case of Ads and ${\cal A}'(a)>0$ for all $a \in (r_{-}, r_{+})$ in the dS case. Of course, one whish to ensure that  ${\cal A}(a)$   is an increasing function on both side of the  throat with spherical topolgy. So   our example corresponds to  a sphere with area  ${\cal A}(a)=4\pi a^{2}$, thus it has  a minimal ``area'' surface reaching  a minimum at the position of  the throat. Now, if we look at the topology of the throat for  $r$ and $\phi$ fixed,  we find that the circular radius function  ${\cal R}(a)=2\pi a$ defines its perimeter  and it can be considered as a less restrictive definition of the wormhole throat  \cite{broni}. 

\subsection{Energy conditions}
There are different, but related energy conditions which can be imposed on the stress-energy tensor \cite{visser1}. Matter satisfying these conditions is denoted ordinary matter, while matter violating these conditions is called exotic matter. 

The \emph{weak energy condition} (WEC) states that for any time-like vector $u^{A}$ it must be $T_{A\,B}u^{A}u^{B}\geq 0$, namely it states that for any observer the measured energy density is non-negative. The WEC also implies, by continuity, the \emph{null energy condition} (NEC), which means that for any null
vector $k^{A}$ it must be $T_{A\,B}k^{A}k^{B}\geq 0$ \cite{visser1}. In an orthonormal basis the WEC reads $\rho\geq 0$, $\rho + P_{l}\geq 0$ $\forall ~ l$  while the NEC takes the form $\rho + P_{l}\geq 0$ $\forall ~ l$.  Besides, the \emph{strong energy condition} states that $\rho + P_{l}\geq 0$ $\forall ~ l$, and $\rho + \sum_{l}P_{l}\geq 0$. If the strong energy condition holds, also the null energy condition holds. However,  the weak energy condition does not follow from the strong energy condition.   

Let us consider the energy-stress tensor for the matter located at wormhole throat (\ref{tem}).  The WEC condition would read $\sigma>0$, $\sigma + P_{\theta}>0$ and $\sigma + P_{\phi}>0$, while NEC only requires  $\sigma + P_{\theta}>0$ and $\sigma + P_{\phi}>0$. Besides, SEC is guaranteed when  $\sigma + P_{\theta}>0$, $\sigma + P_{\phi}>0$ and $\sigma + P_{\theta}+P_{\phi}>0$. In the case of spherically symmetric thin-shell wormholes we have the surface energy density is $\sigma < 0$ and lateral pressures coincide $P_{\theta}=P_{\phi}$. The former fact leads to the violation of the WEC. The reason for such result is  that the flare-out condition must hold; the positivity of the trace condition (\ref{trace}) in turn implies the negativity of $\sigma (a)$.   On the other hand, the sign of $\sigma+P_{\theta}$ or $\sigma+2P_{\theta}$, where $P_{\theta}$  the transverse pressure is not fixed, but it depends on the values of the parameters of the system. Then,  NEC and SEC are not necessarily violated.  We should emphasize that we are not discussing the  energy conditions  associated with the energy-momentum for two exterior regions to the shell ($r>a$), which but the way, it  involves a positive/negative cosmological constant. Such case was examined  by Dias and Lemos \cite{dlemos} within the framework of  d-dimensional general relativity with a cosmological constant for charged thin-shell  wormholes. 
\begin{figure}[!h] \label{fig3}
\begin{center}
\includegraphics[height=7cm, width=8cm]{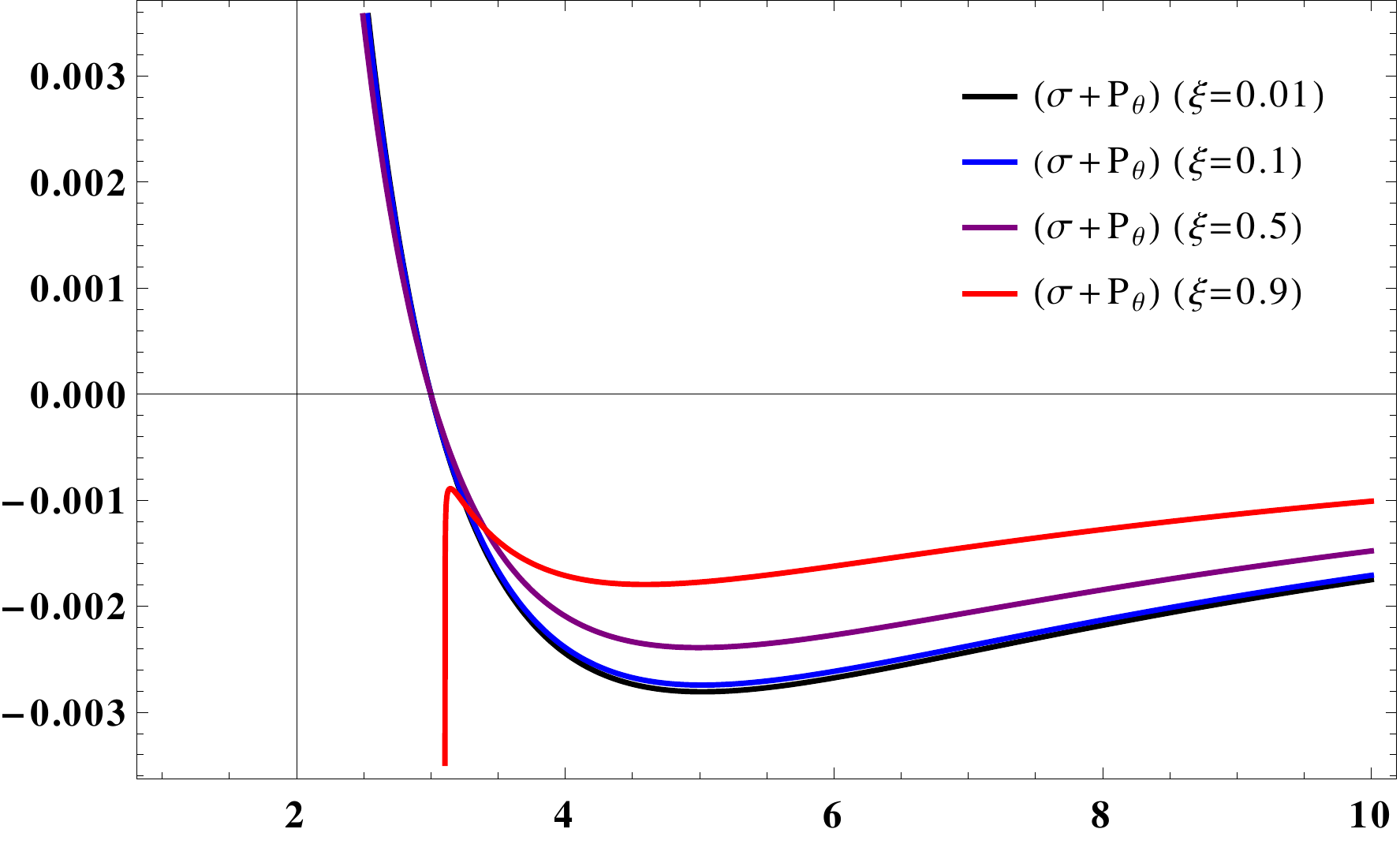}
\includegraphics[height=7cm, width=8cm]{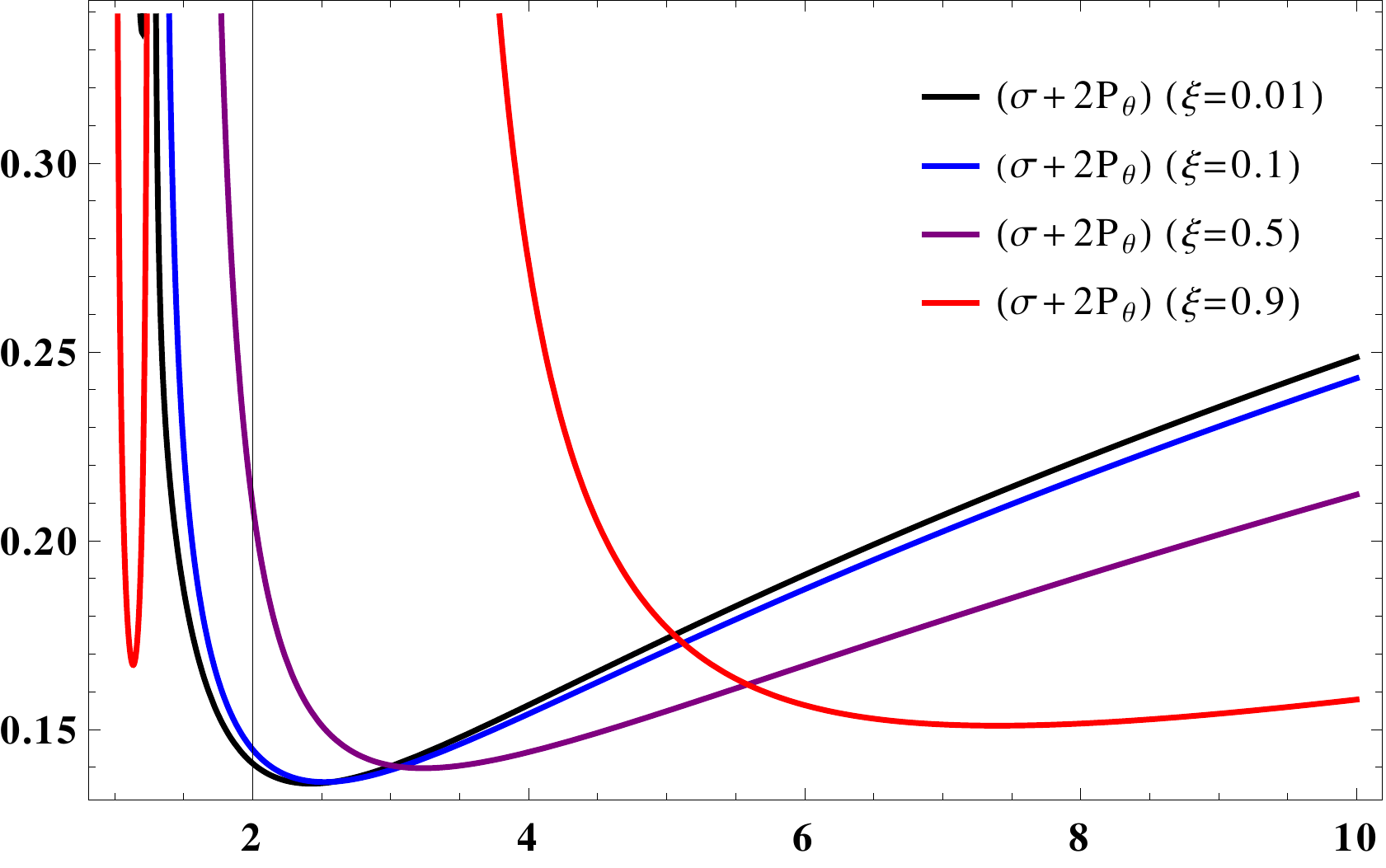}
\caption{Upper panel: It is plotted the condition $\sigma + P_{\theta}$ in terms of the wormhole's radius  for several values of  $\xi \in (0,1)$. Lower panel: It is shown the condition $\sigma + 2P_{\theta}$ in terms of the wormhole's radius  for several values of  $\xi \in (0,1)$. These cases correspond to  acoustic wormholes which are asymptotically Ads. }
\end{center}
\end{figure}

\begin{figure}[!h] \label{fig4}
\begin{center}
\includegraphics[height=7cm, width=8cm]{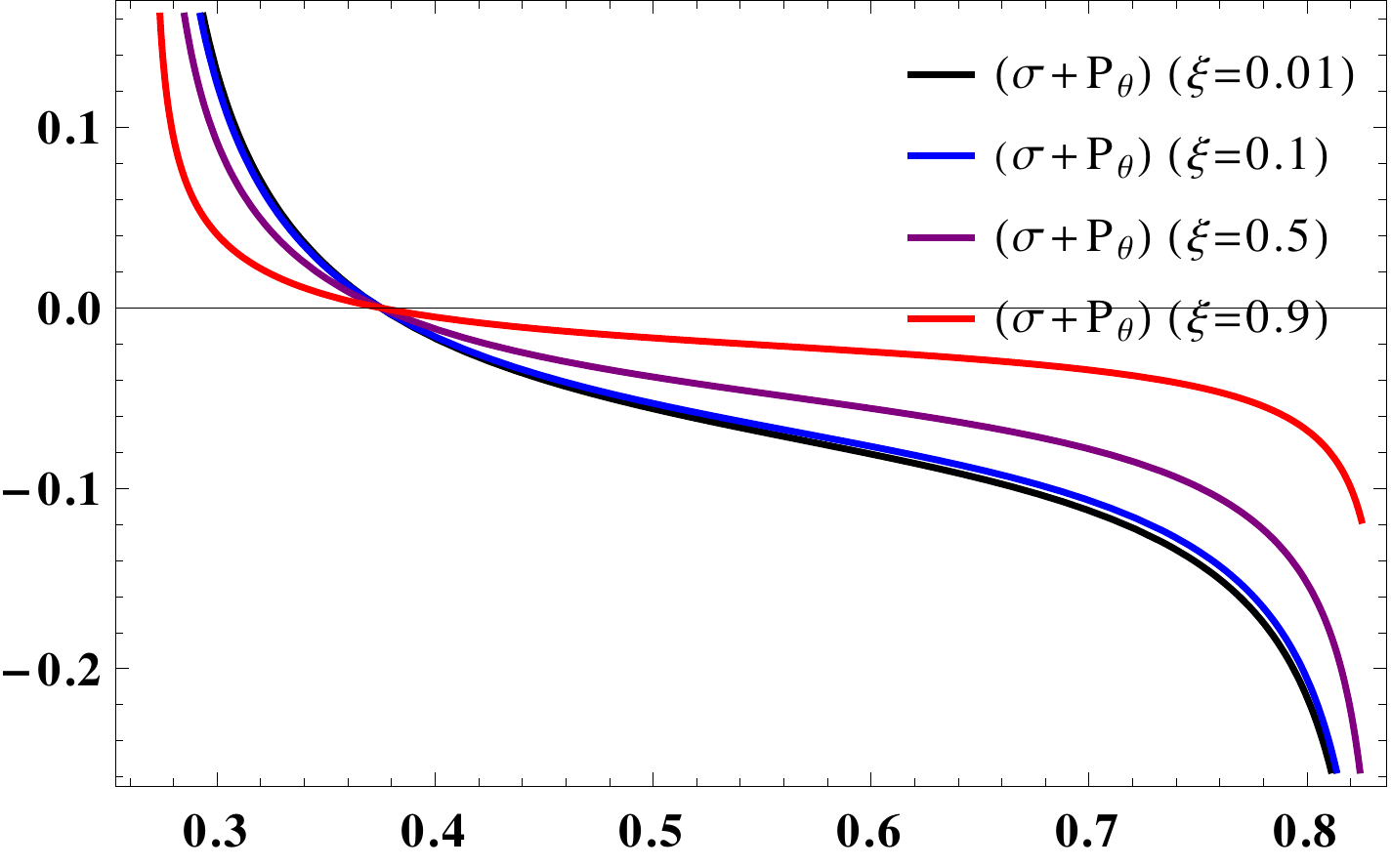}
\includegraphics[height=7cm, width=8cm]{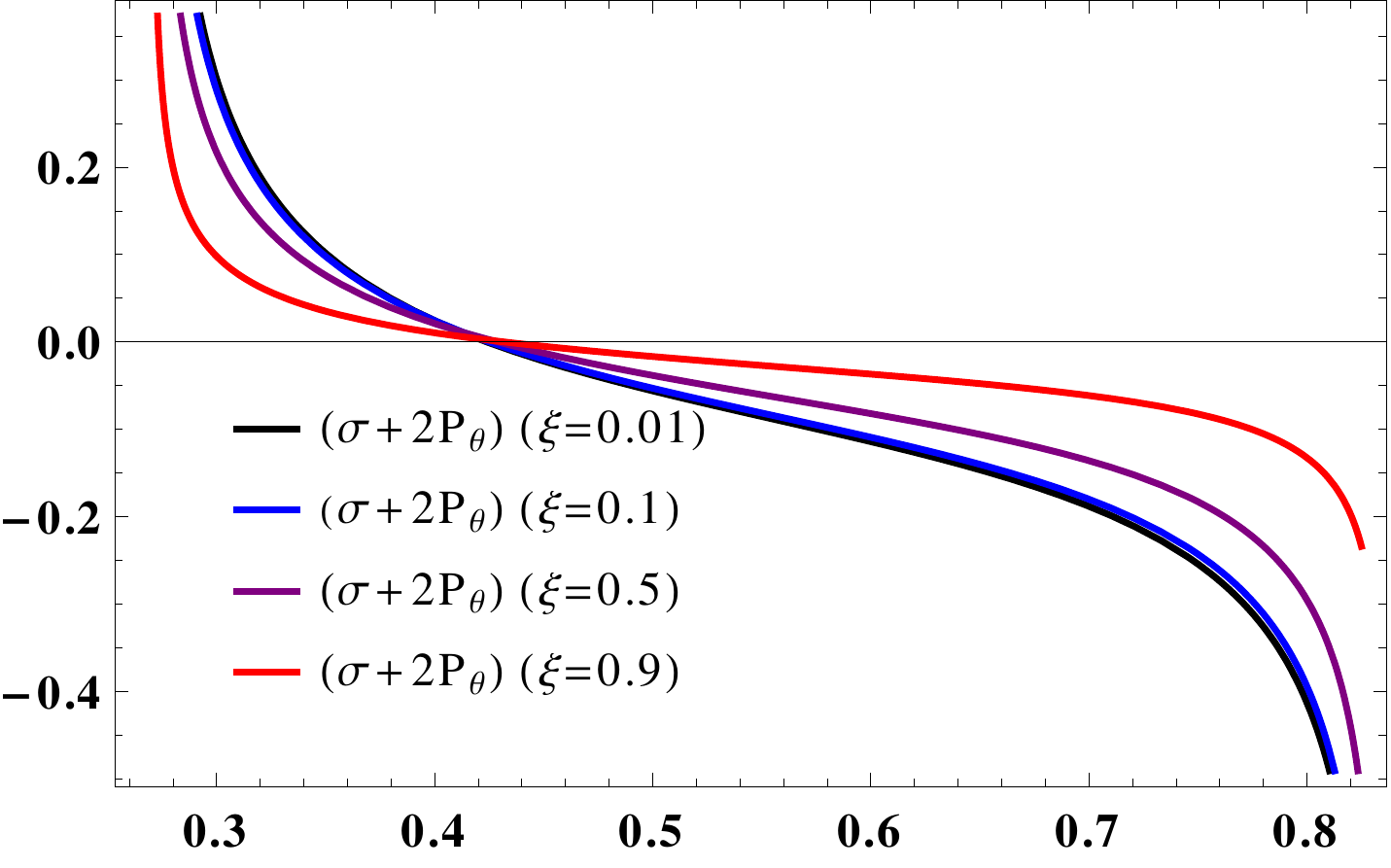}
\caption{Upper panel: The condition $\sigma + P_{\theta}$ in terms of the wormhole's radius  for several values of  $\xi \in (0,1)$. Lower panel: The condition $\sigma + 2P_{\theta}$ in terms of the wormhole's radius  for several values of  $\xi \in (0,1)$.  These cases are associated with  acoustic wormholes which are asymptotically dS.}
\end{center}
\end{figure}

\begin{figure}[!h] \label{fig5}
\begin{center}
\includegraphics[height=7cm, width=8cm]{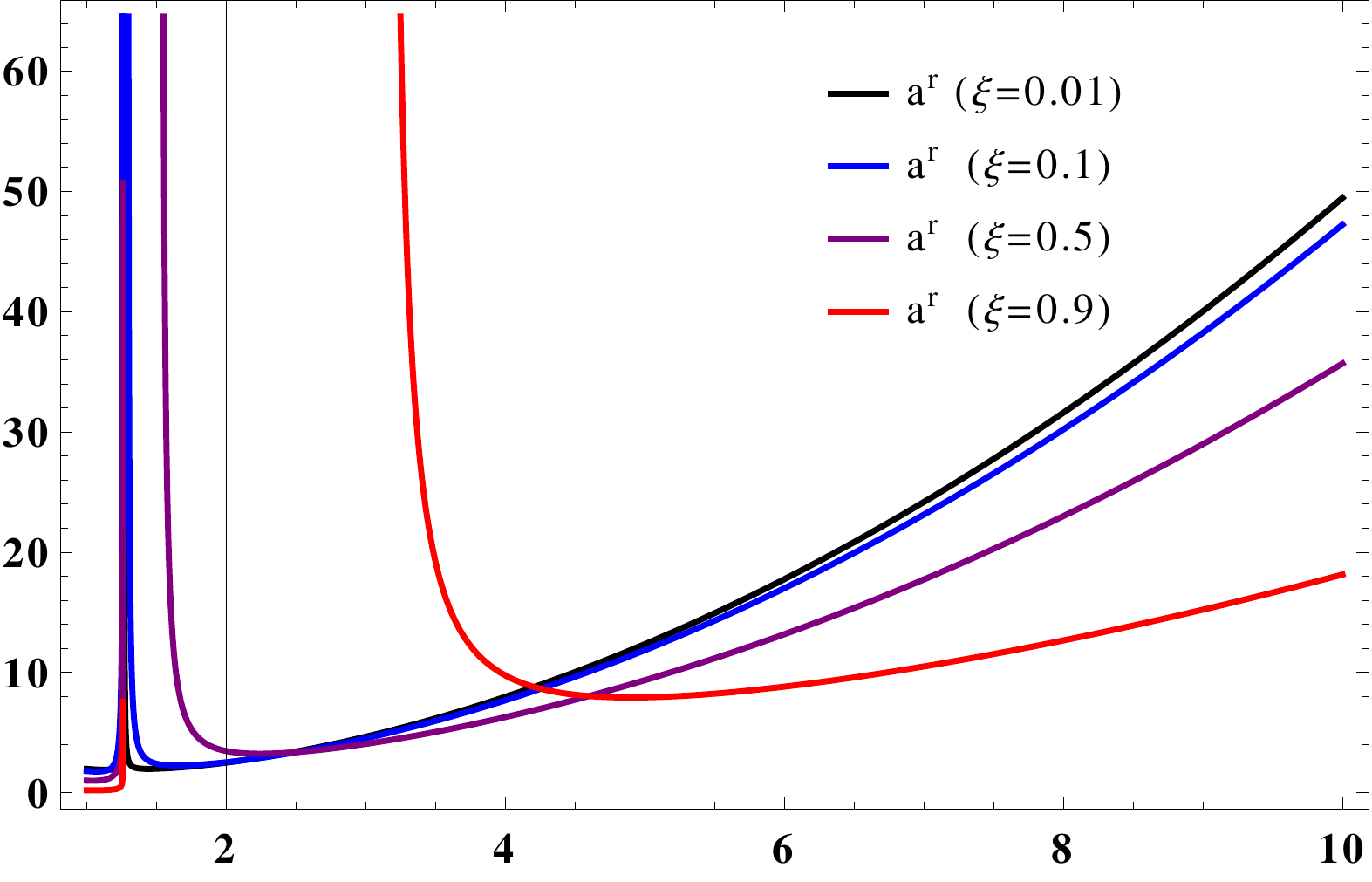}
\includegraphics[height=7cm, width=8cm]{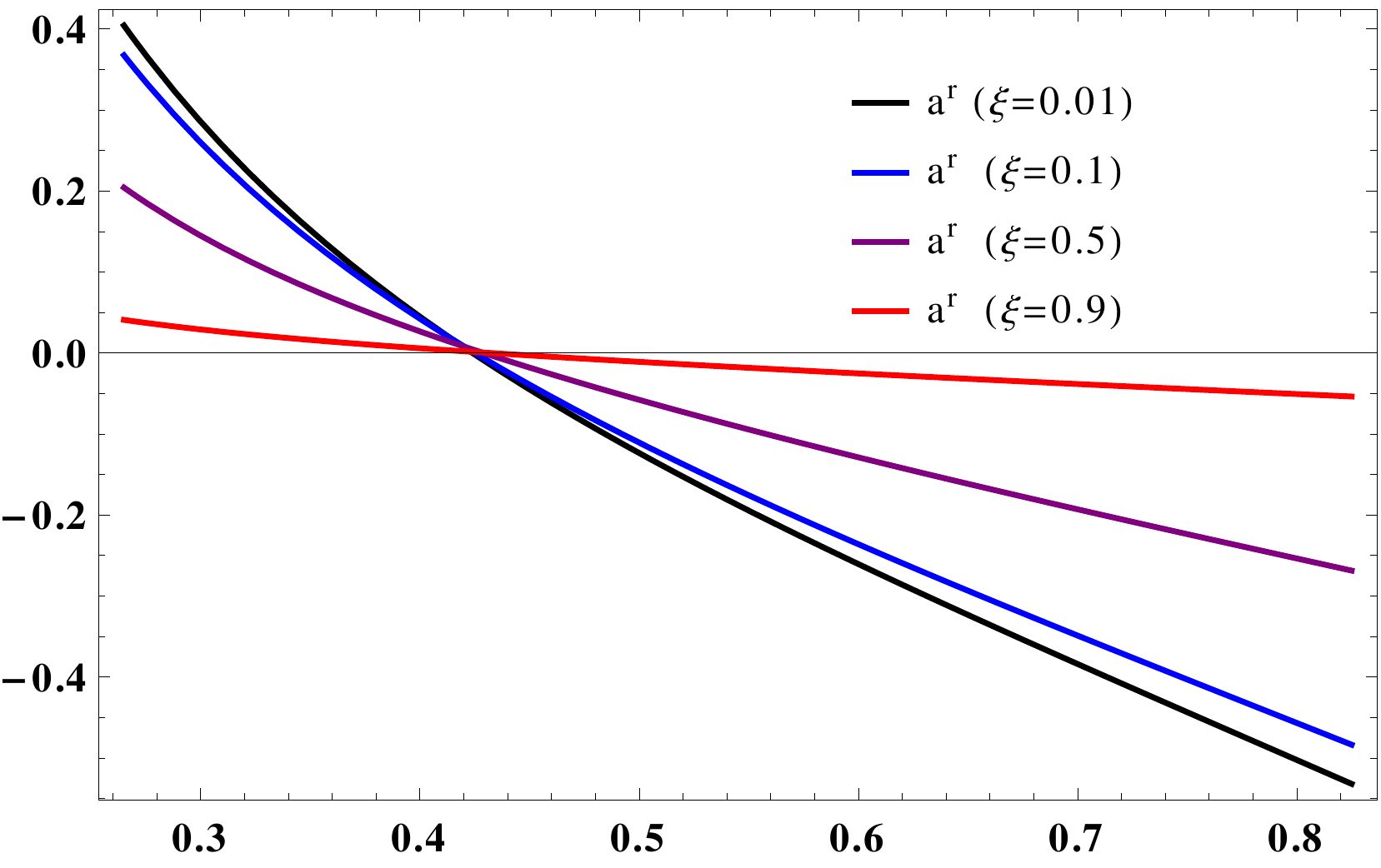}
\caption{Upper panel: Acceleration of the Ads wormhole in terms of its radius  for $\xi \in (0,1)$. Lower panel: Acceleration of the dS wormhole in terms of its radius  for $\xi \in (0,1)$.}
\end{center}
\end{figure}

We must explore whether NEC or SEC are violated or not for static thin-shell acoustic wormholes. To do so, we evaluate  the energy density (\ref{sigma}) and the pressure (\ref{pe}) for $\dot{a}=\ddot{a}=0$.  Then NEC  can be recast as 
 \begin{equation}
 \label{nec1}
\frac{{\cal F}'(a)}{{\cal F}(a)}\geq \frac{{\cal H}'(a)}{{\cal H}(a)},
\end{equation}
while the SEC requires the addition of the following constraint:
 \begin{equation}
 \label{sec1}
\frac{\partial \ln{\cal F}(a)}{\partial a}\geq 0.
\end{equation}
We begin by considering the case of Ads acoustic wormholes with $a \in (r_{+}=1, \infty)$.  We obtain that $\sigma + P_{\theta}$ remains positive for small radius, namely $a \in (1, 3)$,  and takes negative values for $a>3$ [see Fig. 3]. On the other hand,  the constraint $\sigma + 2P_{\theta}$ remains positive for all radius [see Fig. 3]. The previous finding are valid for relativistic acoustic wormhole ($\xi=0.01$) and non-relativistic one ($\xi=0.9$).  Therefore, we can infer that NEC and SEC are satisfied for small radii and they are violated in the case of large radii, regardless of the relativistic nature of the acoustic wormholes. For the acoustic dS wormholes, we obtain that   $\sigma + 2P_{\theta}$ remains positive for $a \in (0.26, 0.37)$ and takes negative values in the interval $a \in (0.37, 0.82)$  while the condition  $\sigma + 2P_{\theta}$ reaches positive values for $a \in (0.26, 0.45)$ and takes negative values in the interval $a \in (0.45, 0.82)$. Hence, NEC and SEC are satisfied when $a \in (0.26, 0.37)$ but they are violated elsewhere [see Fig. 4]. 

\subsection{ Test particles around the acoustic wormhole}
Another appealing attempt to characterize the acoustic wormhole geometries is by looking at the behavior of test particle around them \cite{dlemos}. To do so, we must explore if the test particles are attracted or repelled by static wormhole. In that case, the four velocity is $u^{A}=([\sqrt{{\cal G}(r)}]^{-1},0,0,0)$ then  the four-acceleration takes the form $a^{A}=u^{B}\nabla_{B}u^{A}=\Gamma^{A}_{\eta\eta}[u^{\eta}]^{2}$, being the only non-zero contribution $\Gamma^{r}_{\eta\eta}={\cal F}'/(2{\cal G})$. Then,  a test particle can remain at rest as long as it  keeps a proper acceleration with radial component given by 
 \begin{equation}
 \label{acc}
a^{r}=\frac{{\cal F}'(r)}{2{\cal F}(r){\cal G}(r)}.
\end{equation}
Eq. (\ref{acc}) tells us that acoustic Ads wormholes exhibits an attractive character for all radius while the case of acoustic dS wormholes the story is not so simple. For the latter case, we obtain that it shows an  attractive character for small radii with $a \in (0.26, 0.42)$ but it becomes repulsive in the complementary interval, namely   $a \in (0.42, 0.82)$ [see Fig. 5]. 

\section{Wormhole's stability}
A central aspect of any solution of the equations of gravitation is its mechanical stability. The stability of
wormholes has been thoroughly studied for the case of small perturbations preserving the original symmetry of
the configurations. In particular, Poisson and Visser \cite{11} developed a straightforward approach for analyzing this
aspect for thin-shell wormholes; that is, those which are mathematically constructed by cutting and pasting two
manifolds to obtain a new manifold \cite{12}, \cite{13}. In these wormholes the associated supporting matter is located
on a shell placed at the joining surface; so the theoretical tools for treating them is the Darmois-Israel formal-
ism, which leads to the Lanczos equations \cite{cj0}, \cite{15}. The solution of the Lanczos equations gives the dynamical evolution of the wormhole once an equation of state for the matter on the shell is provided. Such a procedure has been subsequently followed to study the stability of more
general spherically symmetric configurations (see, for example, Refs. \cite{16}-\cite{25}). Moreover, the junction conditions were also used to construct plane symmetric thin-shell wormholes with cosmological constant \cite{26,27}. 

In general to obtain the dynamical picture of the wormholes  can be a very complicated endeavor.
As can be seen from Eqs. (\ref{sigma}-\ref{pe}) the nonlinear character of these expressions makes the idea of obtaining exact solutions very hard to implement. However, we can follow another route and study the stability of static solutions
by linearizing the field equation. A physically interesting wormhole geometry should last enough so that its
traversability makes sense. Thus the stability of a given wormhole configuration becomes a central aspect of its
study. Here we shall analyze the stability under small perturbations preserving the spherical symmetry of the configuration; for this we shall proceed as \cite{11}. As we said, the dynamical evolution is determined by Eqs.  (\ref{sigma}) and (\ref{pe}), or by any of them and the energy-momentum conservation, and to
complete the system we must add an equation of state that relates $p$ with $\sigma$. 

\begin{figure}[!h] \label{fig6}
\begin{center}
\includegraphics[height=7cm, width=8cm]{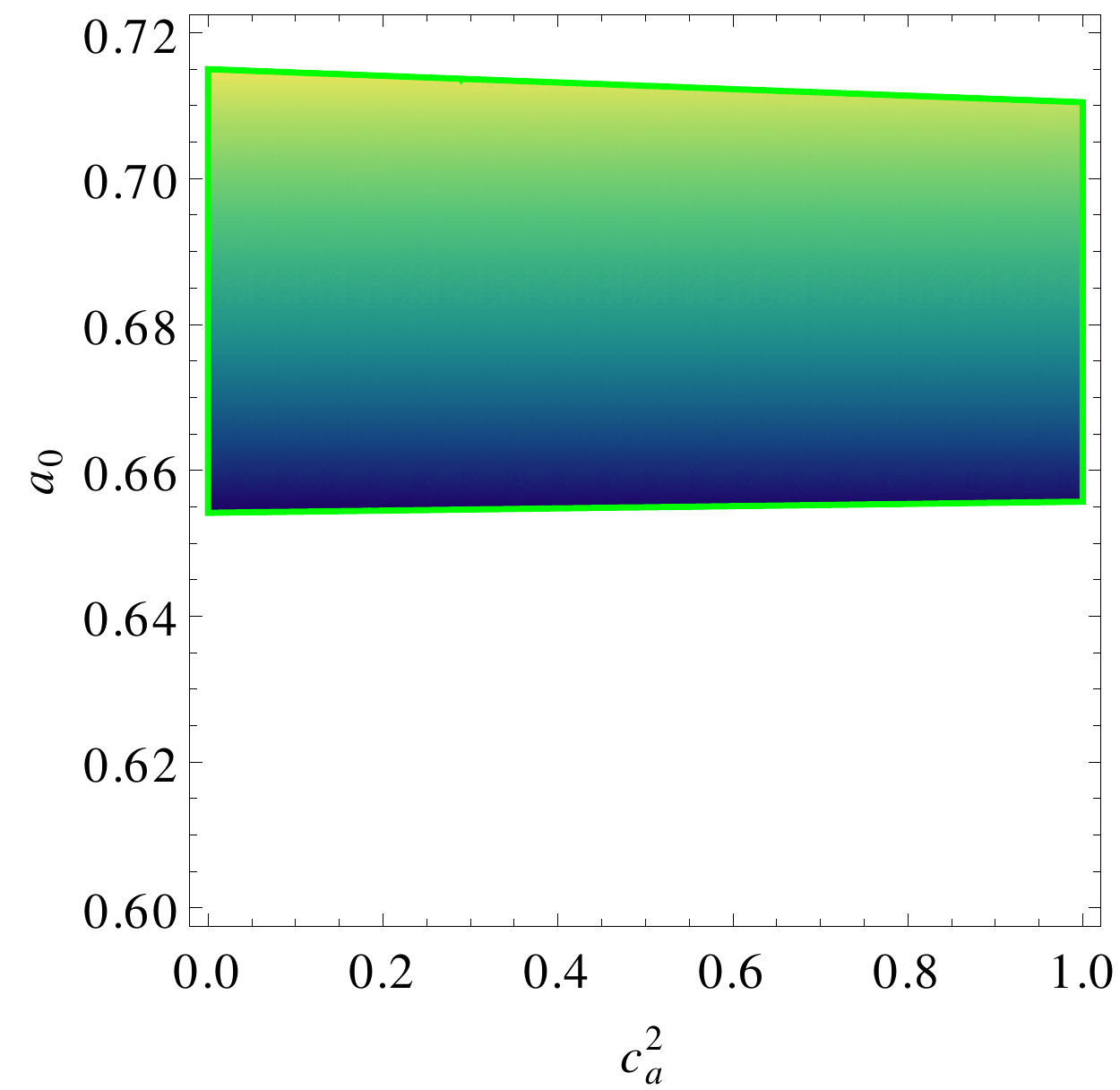}
\includegraphics[height=7cm, width=8cm]{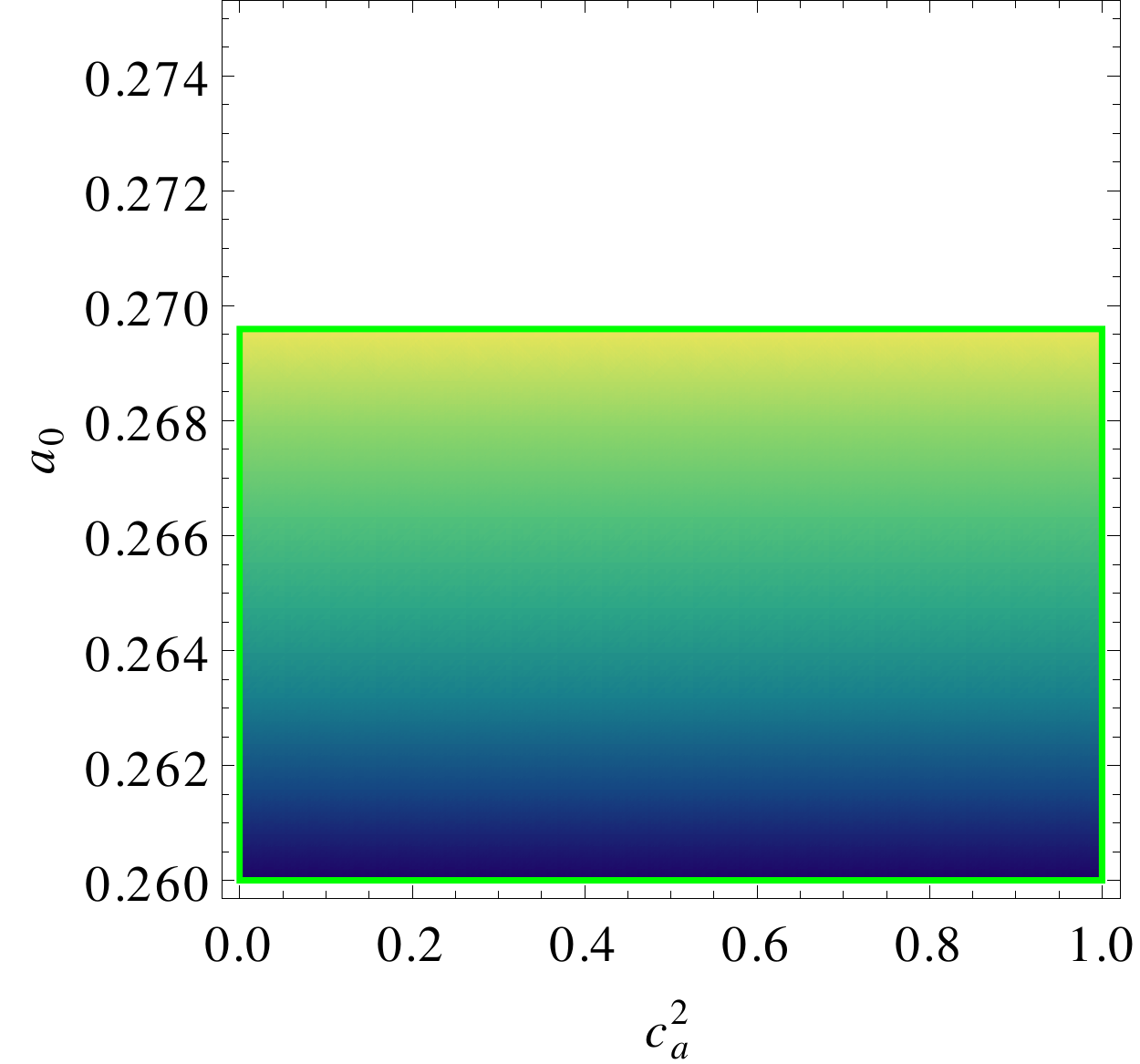}
\caption{Upper panel:Stability regions  in the $a-c^{2}_{a}$ plane for dS wormhole with $\xi=0.01$ in the  large radii case. Lower panel: Stability regions  in the $a-c^{2}_{a}$ plane for dS wormhole with $\xi=0.01$ in the small radii case.}
\end{center}
\end{figure}

\begin{figure}[!h] \label{fig7}
\begin{center}
\includegraphics[height=7cm, width=8cm]{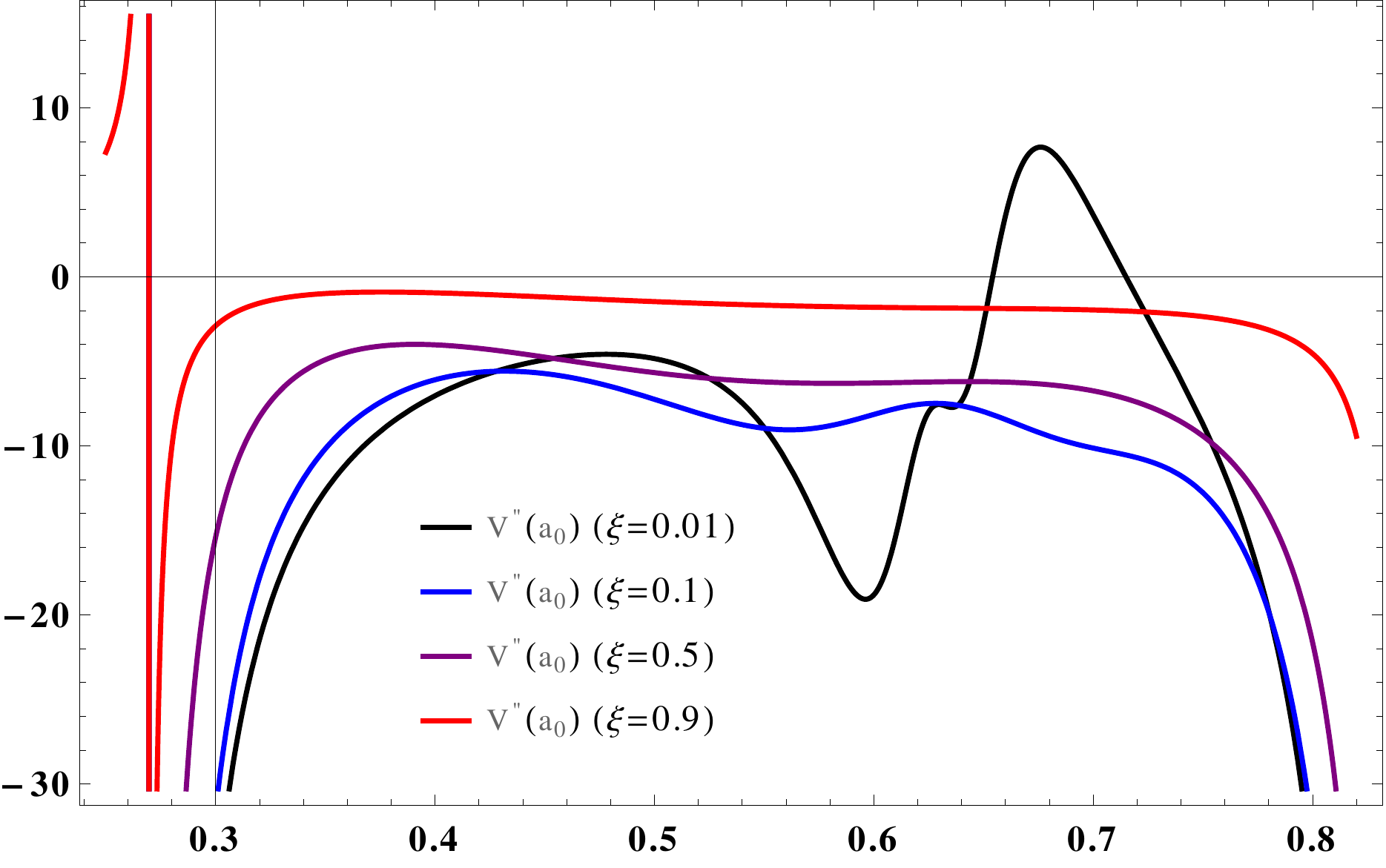}
\includegraphics[height=7cm, width=8cm]{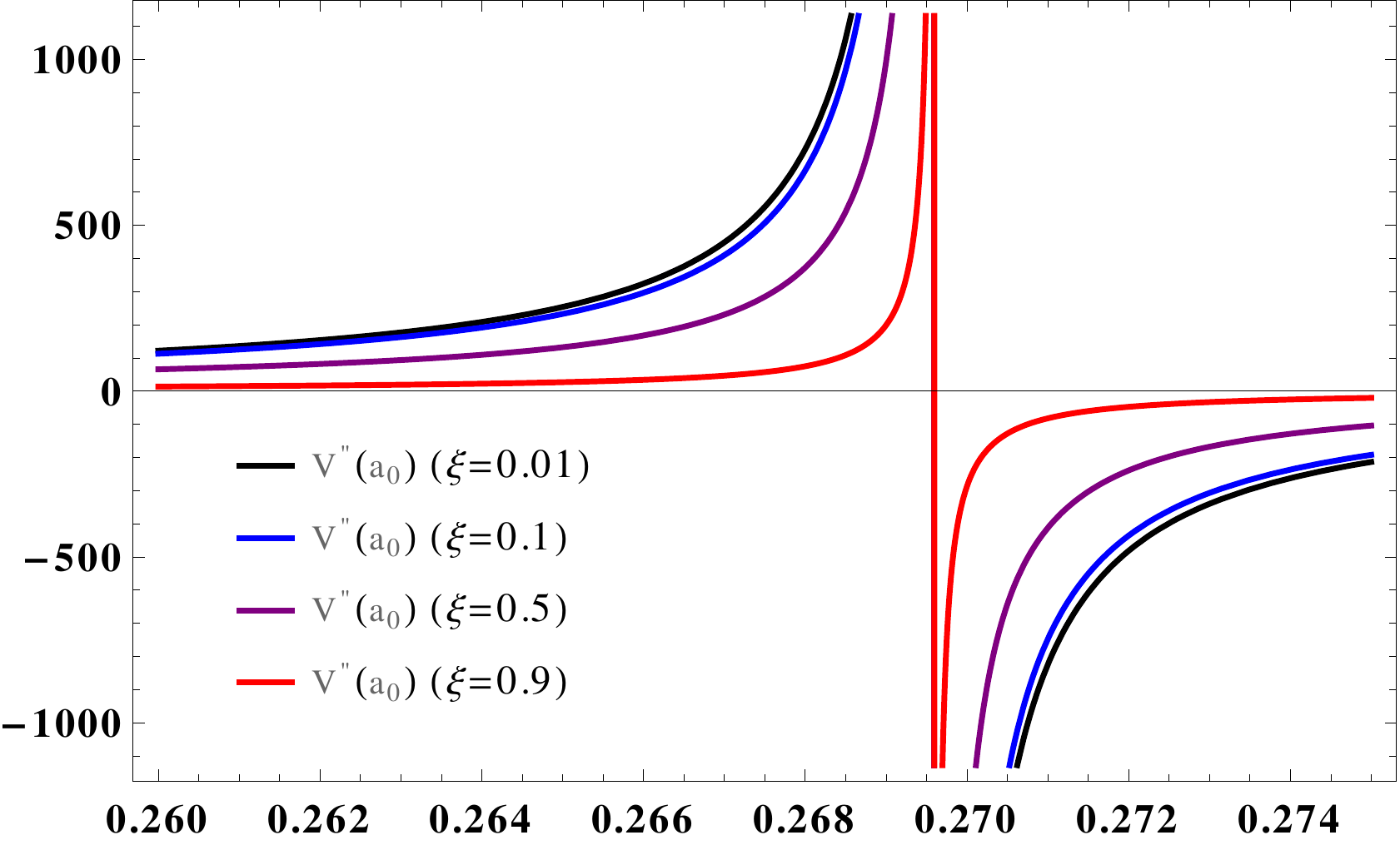}
\caption{Upper panel: Second derivative of the energy potential in terms of the dS wormhole radius for $c^{2}_{a}=0.01$ and $\xi\in  (0,1)$. Lower panel: Second derivative of the energy potential in terms of the dS wormhole radius for $c^{2}_{a}=0.01$ and $\xi\in  (0,1)$; this shows that wormholes with very small radii can be stable.}
\end{center}
\end{figure}

\begin{figure}[!h] \label{fig8}
\begin{center}
\includegraphics[height=7cm, width=8cm]{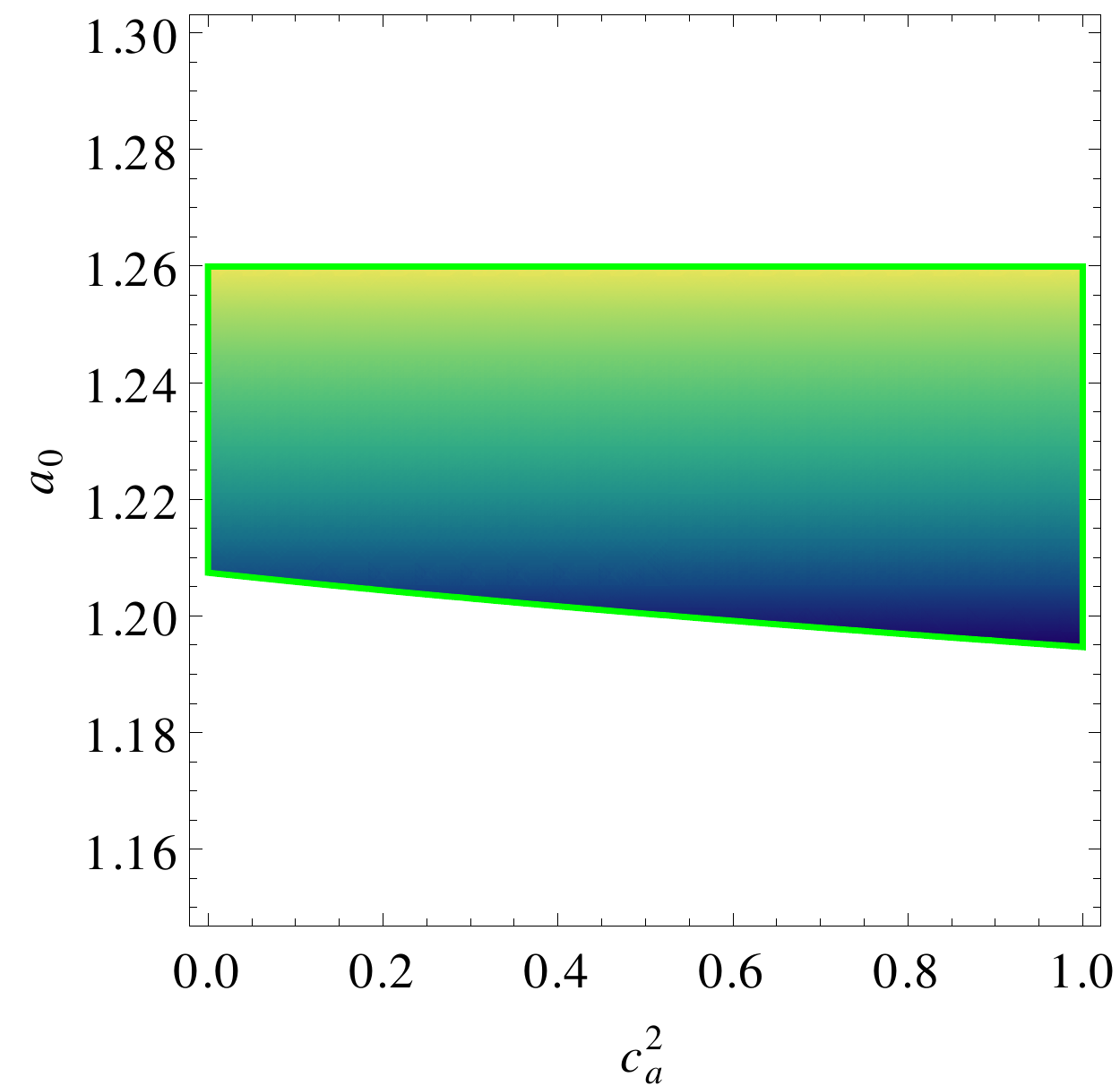}
\includegraphics[height=7cm, width=8cm]{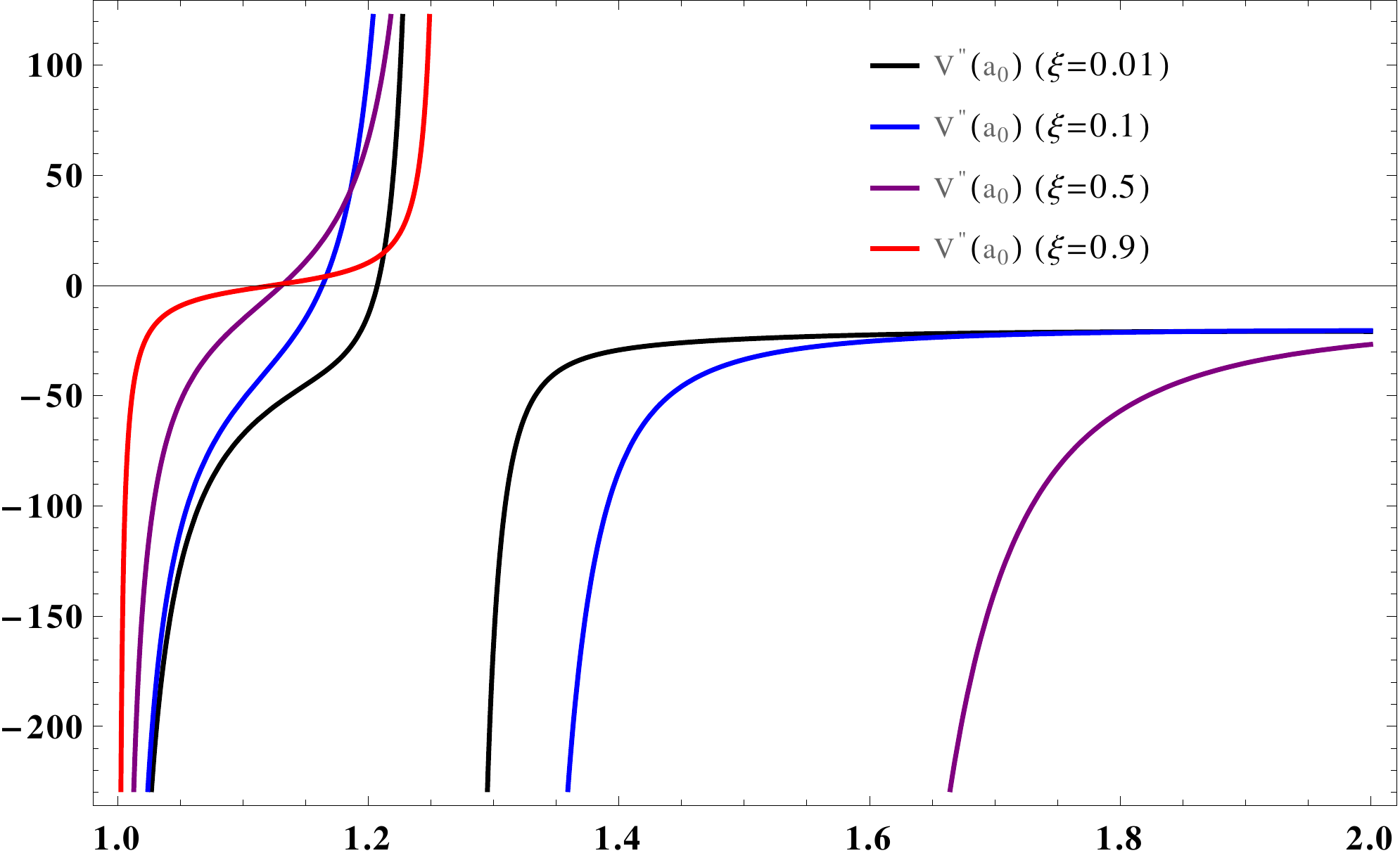}
\caption{Upper panel: Stability regions  in the $a-c^{2}_{a}$ plane for Ads wormhole with $\xi=0.01$. Lower panel: Second derivative of the energy potential in terms of the Ads wormhole radius for $c^{2}_{a}=0.01$ and $\xi\in  (0,1)$; this shows that wormholes with very small radii are allowed to be stable.}
\end{center}
\end{figure}

Our first move to address the stability issue is to recast
Eq. (\ref{sigma}) in such a way that it allows us to get $\dot{a} ={\cal X} (a, \sigma(a))$. Then, by squaring appropriately the energy
density (\ref{sigma}), we obtain the equivalent Hamiltonian constraint of a particle with generalized coordinate $a$ and conjugate momentum $p_{a}=\dot{a}$ in one dimension, namely ${\cal H }= p^{2}_{a}+ V(a, \sigma(a)) \equiv 0$. Such constraint can be written as  

\begin{equation}
\label{masteq}
\dot{a}^{2}=-V[a, \bar{\sigma}],
\end{equation}
with $\bar{\sigma}=4\pi\sigma$. From Eq.(\ref{masteq}) we can write down the potential energy
\begin{equation}
\label{masteq2}
V[a, \bar{\sigma}]= \frac{1}{{\cal G}(a)} -\bar{\sigma}^2 \Big(\frac{{\cal H}}{{\cal H}'}\Big)^2 .
\end{equation}
From the master equation (\ref{masteq2}) we get a single  dynamical equation which completely determines the motion of the wormhole throat after the energy density is selected. To proceed further we need to  make a Taylor expansion of the potential $V$ up to second order around the static solution:
\[V(a)=V(a_{0})+ V'(a_{0})(a-a_{0}) + \frac{1}{2}V''(a_{0})(a-a_{0})^{2}   \]
\begin{equation}
+ {\cal O}[(a-a_{0})^{3}]
\end{equation}
Using (\ref{masteq2}) we get that the first derivative of $V$ is 
\begin{equation}
V'[a_{0}, \bar{\sigma}(a_{0})]=-\frac{{\cal G}'}{{\cal G}^2} -2\bar{\sigma}\frac{{\cal H}}{{\cal H}'}\left[ \bar{\sigma}'\frac{{\cal H}}{{\cal H}'} + \bar{\sigma}\Big(1-    \frac{{\cal H}{\cal H}''}{{\cal H}^{'2}}\Big)\right],  
\end{equation}
while the second derivative of the potential energy  can be written as a superposition of  four different terms, say $V''=\sum^{5}_{I=1}V^{''}_{0I}(a_{0})$. All these terms are evaluated at the static  configurations. Such coefficients are given below:
\begin{equation}
\label{coef1}
V^{''}_{01}= -\frac{{\cal G}''}{{\cal G}^2}+\frac{{\cal G}^{'2}}{{\cal G}^3}, 
\end{equation}
\begin{equation}
\label{coef2}
V^{''}_{02}= 8\bar{\sigma}^2 \frac{{\cal H}}{{\cal H}'}\Big(1- \frac{{\cal H}{\cal H}''}{{\cal H}^{'2}}\Big)\left[\frac{{\cal H}^{'}}{{\cal H}}\Big(1+\frac{\bar{P}}{\bar{\sigma}}\Big) +\frac{\delta \bar{{\cal Q}}}{2}\right] , 
\end{equation}
\begin{equation}
\label{coef3}
V^{''}_{03}= -2\bar{\sigma}^2 \left[1-\frac{{\cal H}^{2}{\cal H}^{'''}}{{\cal H}^{'3}}-\frac{3{\cal H}{\cal H}^{''}}{{\cal H}^{'2}}\Big(1-\frac{3{\cal H}{\cal H}^{''}}{{\cal H}^{'2}}  \Big)\right], 
\end{equation}
\begin{equation}
\label{coef4}
V^{''}_{04}= -2\bar{\sigma}^2 \frac{{\cal H}^{2}}{{\cal H}^{'2}} \left[\frac{{\cal H}^{'}}{{\cal H}}\Big(1+\frac{\bar{P}}{\bar{\sigma}}\Big) +\frac{\delta \bar{{\cal Q}}}{2}\right]^{2}, 
\end{equation}

\[V^{''}_{05}= -2\bar{\sigma}^2 \frac{{\cal H}^{2}}{{\cal H}^{'2}} \left[\frac{{\cal H}^{'}}{{\cal H}}\Big(1+\frac{\bar{P}}{\bar{\sigma}}\Big) +\frac{\delta \bar{{\cal Q}}}{2}\right]\left(\frac{{\cal H}^{'}}{{\cal H}}\big(1+c^{2}_{a}\big) +\frac{\delta \bar{{\cal Q}}}{2}\right)+    \]
\begin{equation}
\label{coef5}
2\bar{\sigma}^2 \frac{{\cal H}^{2}}{{\cal H}^{'2}} \left[\Big(\frac{{\cal H}^{''}}{{\cal H}}-\frac{{\cal H}^{'2}}{{\cal H}^{2}}\Big)\Big(1+\frac{\bar{P}}{\bar{\sigma}}\Big)+\frac{\delta \bar{{\cal Q}'}}{2}\right], 
\end{equation}
where we have used the conservation equation (\ref{tra2}) to express the second derivative $\bar{\sigma}''$ in terms of $\bar{\sigma}'$ and $\bar{\sigma}$. We also defined the adiabatic squared speed sound as $c^{2}_{a}=\bar{P}'/\bar{\sigma}'$, being  $\bar{P}=4\pi P_{\theta}$. Further, we defined the function $\delta \bar{{\cal Q}}=\Big({\cal F}'/{\cal F}+{\cal G}'/{\cal G}+ {\cal H}'/{\cal H} -2{\cal H}''/{\cal H}'\Big)$. To close the system of equations we  postulated that the matter located at the wormhole's throat is described by a linear equation of state which can be parametrized as $\bar{P}=\bar{P}_{0}+ c^{2}_{a}(\bar{\sigma}-\bar{\sigma}_{0})$.  It should be stressed that $c^{2}_{a}$ is a parameter entering the equations of state which is restricted to the interval $(0,1)$ for normal matter.  

It is important to note that wormholes are stable if and only if $V''(a_{0})>0$, while for $V''(a_{0})<0$ perturbations can grow,
 at least until the nonlinear regime is reached. In order to  develop a better understanding of the stability for acoustic wormholes we are going to split our  analysis into parts according to the spatial topology of them; one is associated with Ads wormholes while the other refers to dS wormholes.
This approach will provide us with some interesting insights on how the topology of acoustic wormhole can affect the linear stability of them, and therefore it will be  useful for later comparison.
 
We begin by considering the case of acoustic dS wormholes.  For $L=1$ and $r_{0}=1/4$ the wormhole has finite radius $a_{0} \in (0.26, 0.82)$. Fig. (6) shows that there are two small patches in the $a_{0}-c^{2}_{s}$ plane where in each patch is possible to find static configurations which remain stable under radial perturbations. For $\xi=0.01$ (ultra-relativistic BEC), wormholes are stable as along as their radii belong to the interval $(0.26, 0.27)\cup (0.66, 0.71)$, otherwise they are unstable ones. Interestingly enough, there is a nice correlation between the stability zones and the non-violation of NEC and SEC. As a result, we obtain that  stable dS wormholes with  small radii $a_{0} \in(0.26, 0.27)$ are accommodated as static configurations which do not violate NEC and SEC. We also confirm our previous analysis by plotting the behavior of $V''(a_{0})$ for different values of $\xi$ but fixing  the adiabatic speed sound parameter as $c_{a}=\sqrt{0.01}$ [see Fig. (7)]. As it can be seen only wormholes with ultra-relativistic BEC parameter $\xi=0.01$ are stable at large radii.  Numerical analysis shows up that  only are allowed wormholes with very small radius, namely  $a_{0} \in (0.26, 0.27)$, when  the BEC parameter approaches to  non-relativistic values, for instance $\xi=0.9$.

We carry on our analysis by exploring the case of acoustic Ads wormholes. For practical purposes we fix $L=1$ and $r_{0}=2$, so the original manifold exhibits an horizon at $r_{+}=1$. Fig. (8) shows that only wormholes with small radii are allowed to be stable for $\xi=0.01$. The stability zone is  $a_{0} \times c^{2}_{a}=(1.21, 1.26) \times (0,1)$. Further, higher values of the BEC parameter $\xi$ does not change such findings. Our conclusion is confirmed by plotting the $V''(a_{0})$ for different values of $\xi$. As a result, we find that all acoustic wormholes asymptotically Ads are unstable under radial perturbation regardless the value taken by  the BEC parameter. Once again, the correlation between stability regions and the violation of NEC and SEC conditions is established. Stable Ads wormhole satisfy both NEC and SEC, whereas unstable Ads wormhole with large radii violate both energy conditions.

\section{Summary}

We have considered the construction of acoustic thin-shell wormholes asymptotically Ads/dS where the original manifold emerges within the context of relativistic Bose-Einstein condensates. Our previous analysis was possible provided  it has been shown  that black holes metric are connected with acoustic effective geometry associated with BEC up to a conformal factor.  

We have shown that dS/Ads acoustic wormholes both satisfied the (trace) flare-out condition, namely ${\rm Tr}({\cal K}_{ij})>0$ , which is equivalent to state that the static configurations are supported by negative energy density. Nevertheless,  we have shown that if the aforesaid condition cannot be applied the transversavility of these configurations can be guaranteed provided the areal condition ${\cal A}'(\Sigma_{\tau})>0$ along with the perimeter condition  ${\cal P}'(\Sigma_{\tau})>0$ hold. To further characterize these wormholes, we explored the behavior of a test particle near the wormhole's throat. Interestingly enough, we found that  acoustic Ads wormholes exhibit an attractive character for all radius while  acoustic dS wormholes only are  attractive as long as their radii remain very small otherwise they repel the test article.

One of the main reasons to explore the existence of thin-shell acoustic wormholes is that one would be interested in those which  require a minimal amount of exotic matter. At least, one could be interested in studying the possibility of having configurations which do not necessarily violate all the energy conditions. For the latter reason, we have explored the violation or not of the energy conditions and their connections with the stability of acoustic wormholes. To be more precise, we  have found that static dS wormholes can remain stable under radial perturbations as long as they have small radii. For instance, we  took $\xi=0.01$ (ultra-relativistic BEC parameter) and obtained  stable wormholes with radii $ \in (0.26, 0.27)\cup (0.66, 0.71)$, otherwise they are unstable ones. Interestingly enough, those  stable dS wormholes with  small radii $a_{0} \in(0.26, 0.27)$ do not violate NEC and SEC. In addition, we have explored  the stability of  acoustic Ads wormholes. We noted that wormholes with small radii are allowed to be stable for ultra-relativistic BEC parameter ($\xi=0.01$). Nevertheless, we arrived at the conclusion that moving toward the non-relativistic regime (with a higher BEC parameter)  does not change such findings. Furthermore,  we showed that acoustic Ads wormholes with large radius are unstable under radial perturbation regardless the value taken by  the BEC parameter. In addition,  we pointed out that stable Ads wormhole satisfy both NEC and SEC but unstable Ads wormhole with large radii violate both energy conditions.

\section{Acknowledgements}
 M. G. R  and J.P. M.G are  supported by Conselho Nacional de Desenvolvimento Cient\'ifico e Tecnol\'ogico (CNPq)- Brazil.
The work of H. Moradpour has been supported financially by Research
Institute for Astronomy \& Astrophysics of Maragha (RIAAM) under
project No.1/4165-1. The work of A.\"{O} is supported by the Chilean FONDECYT Grant No. 3170035. 

\end{document}